\documentclass[12pt,preprint]{aastex}

\shorttitle{Spectro-interferometry and radiative transfer modeling of the
  accretion disk around MWC~147}
\shortauthors{Kraus, Preibisch \& Ohnaka}

\begin{document}

\title{
Detection of an inner gaseous component in a Herbig~Be star accretion
  disk: 
Near- and mid-infrared spectro-interferometry and 
radiative transfer modeling of \object{MWC~147}\altaffilmark{1}
}

\altaffiltext{1}{Based on observations made with ESO telescopes at the La Silla
  Paranal Observatory under programme IDs 074.C-0181, 076.C-0138, and
  078.C-0129. In addition, this work is based in part on archival data
  obtained with the Spitzer Space Telescope, which is operated by the Jet
  Propulsion Laboratory, California Institute of Technology under a contract
  with NASA. }


\author{Stefan Kraus, Thomas Preibisch and Keiichi Ohnaka}
\affil{Max Planck Institut f\"ur Radioastronomie, Auf dem H\"ugel 69, 53121 Bonn, Germany}
\email{skraus@mpifr-bonn.mpg.de}




\begin{abstract}
We study the geometry and the physical conditions in the inner
(AU-scale) circumstellar region around the young Herbig~Be star 
MWC~147 using long-baseline spectro-interferometry in the near-infrared (NIR
$K$-band, VLTI/AMBER observations and PTI archive data) 
as well as the mid-infrared (MIR $N$-band, VLTI/MIDI observations).
The emission from MWC~147 is clearly resolved and has a characteristic
physical size of $\sim1.3$~AU and $\sim9$~AU at $2.2~\mu$m and $11~\mu$m
respectively (Gaussian diameter). 
The MIR emission reveals asymmetry consistent with a disk structure
seen under intermediate inclination. The spectrally dispersed AMBER and MIDI
interferograms both show a strong increase in the characteristic size
towards longer wavelengths, much steeper than predicted by analytic disk
models assuming power-law radial temperature distributions.

We model the interferometric data and the spectral energy distribution
of MWC~147 with 2-D, frequency-dependent radiation transfer simulations.
This analysis shows that models of spherical envelopes or
passive irradiated Keplerian disks (with vertical or curved puffed-up inner 
rim) can easily fit the SED, but predict much lower visibilities
than observed; the angular size predicted by such models is 2 to 4 
times larger than the size derived from the interferometric data, 
so these models can clearly be ruled out.
Models of a Keplerian disk with optically thick gas emission from an active
gaseous disk (inside the dust sublimation zone),
however, yield a good fit of the SED and simultaneously reproduce the absolute
level and the spectral dependence of the NIR and MIR visibilities.
We conclude that the NIR continuum emission from MWC~147 is 
dominated by accretion luminosity emerging from an optically thick 
inner gaseous disk, while the MIR emission also contains contributions from
the outer, irradiated dust disk.
\end{abstract}




\section{Introduction} \label{sec:intro}

The spatial structure of the circumstellar material around Herbig~Ae/Be (HAeBe)
stars, i.e.\ intermediate-mass, pre-main sequence stars, is still a matter of
debate.  Until recently, the spatial scales of the inner circumstellar
environment (a few AU, corresponding to $\lesssim $0\farcs1) were not
accessible to optical and infrared imaging observations, and conclusions drawn
on the spatial distribution of the circumstellar material
were, in most cases, entirely based on the modeling of the 
spectral energy distribution (SED).
However, fits to the observed SEDs are highly ambiguous,
and very different models such as 
spherical envelopes \citep{mir97}, geometrically thin disks \citep{hil92},
disks surrounded by a spherical envelope \citep{nat95, nat01, mir99},
flared outer disks, puffed-up inner disk rims \citep{dul01}, and 
disk plus inner halo models \citep{vin06}
have been used to successfully fit the observed SEDs of HAeBes.
It was also unclear how the accretion of circumstellar material onto
the star affects the structure and emission of the
inner circumstellar environment.

In the last couple of years, long-baseline interferometry in the
near- (NIR) and mid-infared (MIR) spectral range provided, for the first time, 
direct spatial information on the {\em inner} circumstellar regions of
young stars at scales of $\lesssim 10$~milli-arcseconds 
\citep[mas; e.g.,][]{mal98, mil99, ake00, mil01, wil03, eis04, lei04, mon05,mal07}.
The interferometric data were usually interpreted by
fitting simple analytic models for the brightness distribution to
the measured visibilities. The characteristic sizes of the emitting regions 
derived in this way were found to be correlated to the stellar luminosity,
consistent with the idea that the NIR continuum emission mainly
traces hot dust close to the dust sublimation radius,
i.e.~where the grains are heated above their sublimation temperature
($T_{\rm sub} \sim 1\,500$~K) and destroyed by the (stellar) radiation field
\citep{tut02, mon02}.
While the expected simple $R_{\star}\propto L_{\star}^{1/2}$ scaling law
between stellar luminosity $L_{\star}$ and NIR size $R_{\star}$ appears to
hold throughout the low- to medium-luminosity part of the observed stellar
sample, some very luminous early B-type stars exhibited considerably smaller 
NIR sizes than predicted by this relation \citep{mon02,eis04,mon05}.
\citet{mon02} suggested that this might be due to the presence of an inner
gaseous disk, which shields the dust disk from the strong stellar ultraviolet
(UV) radiation. Since this shielding would be most efficient for hot stars, it
would allow the inner rim of the dust disk around B-type stars to exist closer
to the star. Several subsequent studies favour ``classical'' accretion disk
models \citep{eis04,mon05,vin07} or ``two-ring'' models 
\citep{eis07}, in which the infrared emission contains contributions
from the thermal emission of optically thick gas in the innermost disk
regions. 

Following initial attempts by \citet{hin01}, who used
single-dish nulling interferometry to put upper limits of $\lesssim 20$~AU
on the MIR size of some Herbig~Ae stars, a large sample of HAeBes disks
could be resolved with the VLTI/MIDI long-baseline interferometer at MIR
wavelengths.
In contrast to the NIR flux, which originates from hot 
dust in the innermost disk regions close to the dust sublimation 
radius, the MIR emission also traces cooler
($\sim 200-300$~K) dust further out in the disk. 
\citet{lei04} determined characteristic dimensions of the $10~\mu$m emitting 
regions for a sample of HAeBes, which ranged from 1~AU to 10~AU.

Due to the quite limited $uv$-plane coverage of most existing 
infrared interferometric data sets, most published studies were only able
to derive estimates of characteristic sizes and, in some cases,
to look for possible elongation of the emitting region, but
could not investigate the geometry of individual sources in detail.
A more
comprehensive  interferometric study of one Herbig~Ae star, HR~5999, was 
recently performed by our group \citep{pre06}. Based on a set of ten MIDI 
measurements at different projected baseline lengths and position angles (PAs),
modeling with 2-D frequency-dependent radiation transfer 
simulations provided relatively detailed information on the disk size and
inclination.\newline

In this study, we combine, for the first time, NIR and MIR interferometric
observations to constrain the spatial structure of the dust and gas
environment around a Herbig~Be star.
Besides the wide spectral coverage (from $\sim 2$ to $13~\mu$m), our
interferometric data is also dispersed into several spectral channels
(resolution $R=\lambda/\Delta\lambda \approx 30$), allowing us to measure the
wavelength-dependence of the visibility over the $K$- and $N$-band and within
spectral features such as the silicate emission feature around $10~\mu$m.
In combination with detailed 2-D radiative transfer modeling, this
spectro-interferometric data set provides unique information about the 
inner circumstellar structures of this star.

\medskip

\begin{deluxetable}{rcl}
\tabletypesize{\scriptsize}
\tablecaption{Stellar parameters for MWC~147 from \citet{her04b}, assuming
   $R_V=3.1$.
  \label{tab:stellarproperties}}
\tablewidth{0pt}
\centering

\tablehead{
  \colhead{Parameter} & \colhead{} & \colhead{Value}
}
\startdata
  Spectral Type    &             & B6\\
  Effective Temperature & $T_{\mathrm{eff}}$ & 14\,125~K\\
  Luminosity       & $L_{\star}$ & 1\,550~$L_{\sun}$\\
  Mass             & $M_{\star}$ & 6.6~$M_{\sun}$\\
  Age              & $t$         & 0.32~Myr\\
  Distance         & $d$         & 800~pc\\
  Extinction       & $A_V$       & 1.2~mag\\
  Stellar Radius   & $R_{\star}$ & 6.63~$R_{\sun}$\\
\enddata
\end{deluxetable}

MWC~147 (alias HD\,259431, BD+10\,1172, HBC 529, V700\,Mon)
is  a well-studied Herbig~Be star in Monoceros.  
There is some uncertainty concerning
the distance and the physical parameters of this star.
From the analysis of the Hipparcos parallax data by \citet{van98}, a lower
limit on the distance of $>\!130$~pc was derived, while a reanalysis suggested
a distance of $290^{+200}_{-84}$~pc \citep{ber99}. This distance estimate,
however, is in conflict with the apparent location of
MWC~147 in the NGC\,2247 dark cloud, which is a part of the
cloud complex in the Monoceros OB1 association at a distance of $\sim
800-900$~pc \citep{oli96}.
We assume a distance of 800~pc for MWC~147 (consistent with
most other recent studies) and use the main stellar parameters as
listed in Tab.~\ref{tab:stellarproperties}, which were taken from 
\citet{her04b}.

Numerous observational results strongly suggest the presence of a
circumstellar disk around MWC~147. The object shows a strong infrared
excess of about 6~mag at MIR wavelengths, clearly demonstrating the
presence of circumstellar material.
\citet{hil92} fitted the SED of MWC~147 with a model assuming an 
accretion disk and estimated an accretion rate of
$\dot{M}_{\rm acc} = 1.01 \times 10^{-5}~M_{\sun} $yr$^{-1}$.
MIR (10~$\mu$m and 18~$\mu$m) imaging observations 
revealed an elongated diffuse emission component around MWC~147 along PA
$\sim$50\degr, extending out to $\sim$6\arcsec\ and contributing
$\sim 34$\% to the total flux \citep{pol02}. \citet{man94} determined the
1.1~mm flux of MWC~147 and estimated the mass in the circumstellar
disk/envelope to be $< 0.09~M_{\sun}$. 
The study of the far-UV spectrum of MWC~147 by \citet{bou03}
also suggested the presence of a flared circumstellar 
disk.
\citet{pol02} imaged MWC~147 in the MIR and concluded that the star is
surrounded by a moderately flared disk and probably an extended envelope.
Measurements by \citet{jai90} showed a significant amount of linear polarization
($\sim$1\% along PA $\sim 106\degr$) but no wavelength-dependence of the
polarization.
The high observed rotational velocity of $v \sin i = 90$~km\,s$^{-1}$
\citep{boe95} suggests a high inclination of the rotation axis of MWC~147 with
respect to the line-of-sight. This implies that the orientation of the
circumstellar disk should be closer to edge-on than to face-on.

Recently, \citet{bri07} presented a high-resolution NIR spectrum of MWC~147, 
showing a strong Br$\gamma$ emission line.  Using an empirical relation 
between the Br$\gamma$ luminosity and the accretion rate \citep[as derived
from UV veiling;~][]{van05}, 
they derive a mass accretion rate of $\dot{M}_{\rm acc} = 4.1 \times
10^{-7}~M_{\sun} $yr$^{-1}$.  As discussed in their Section 5.2, it
is well possible that this method underestimates the true mass accretion
rate.  Adopting the stellar parameters used in our study
(Tab.~\ref{tab:stellarproperties}) will also result in a larger value for the
derived accretion rate.

Evidence for a strong stellar wind from MWC~147 comes from the observed
P~Cygni profiles in several emission lines \citep{bou03}. 
A quantitative modeling of FUSE spectra revealed multiple absorption
components with different temperatures, consistent with a flared disk
interpretation \citep{bou03}.
Based on the intensity ratio of infrared hydrogen lines, \citet{nis95}
estimated a mass loss rate of $2.0 \pm 0.4 \times 10^{-7}~M_{\sun}$yr$^{-1}$,
which is slightly higher than the mass loss rates determined from radio
observations \citep[$0.68 \times 10^{-7}~M_{\sun}$yr$^{-1}$, ][]{ski93}.

The star has a faint visual companion at a projected separation of
3\farcs1~\citep[$\sim 2\,500$~AU, $\Delta R=6.82$, ][]{bai06}.
While \citet{vie94} classified MWC~147 as a spectroscopic binary with a
period of about one year, this claim could not be confirmed in more recent
observations \citep{cor99}.  

First interferometric measurements on MWC~147 were presented by
\citet{mil01}, providing an upper limit on the $H$-band size.
\citet{ake00} observed MWC~147 with the Palomar Testbed Interferometer (PTI)
and resolved its emission in the $K$-band at baselines around 100~m. 
They derived a best-fit Gaussian FWHM diameter of
1.38~mas (=1.1~AU) in the $K$-band.

\section{Observations and data reduction}


\begin{deluxetable}{lccccrrll} \rotate
\tabletypesize{\scriptsize}
\tablecaption{Observation log for interferometric observations on MWC~147.  For more
  detailed information about the calibrator stars, we refer to
  Tab.~\ref{tab:calibrators}.
  \label{tab:observations}}
\tablewidth{0pt}

\tablehead{
  \colhead{Instrument} & \colhead{Date}  & \colhead{HA}  & \colhead{Band/}  & \colhead{Baseline}  &  \multicolumn{2}{c}{Projected Baseline} &  \colhead{Calibrators}  & \colhead{Ref.}\\
  \colhead{}           & \colhead{(UT)}  & \colhead{}    & \colhead{Spectral Mode}  & \colhead{}  &  \colhead{Length [m]} &  \colhead{PA [$^\circ$]}  & \colhead{}  & \colhead{}
}
\startdata
  \multicolumn{9}{c}{\textbf{Near-Infrared}}\\
  \noalign{\smallskip}
  \tableline
  \noalign{\smallskip}
  IOTA/FLOUR        & 1998        &            & H        & 38~m           & $\sim 22$  &  $\sim 25$                & see Ref.                  & (1)\\
  IOTA/FLOUR        & 1998        &            & K'       & 38~m           & $\sim 21$  &  $\sim 15$                & see Ref.                  & (1)\\
  PTI               & 1999, 2000, 2003 &  $<0$ & K        & NS            & 105.1      &  29                       & HD\,42807, HD\,43042,     & (2), (3)\\
                    &             &  $\geq0$   &          & NS            & 98.9       &  17                       & HD\,43587, HD\,46709,     &\\
                    &             &            &          &               &            &                           & HD\,50692                 &\\
  PTI               & 2004        &            & K        & NW            & 85.7       &  76                       & HD\,43042, HD\,46709      & (3)\\
                    &             &            &          &               &            &                           & HD\,50692                 &\\
  PTI               & 2003, 2004  & $<0$       & K        & SW            & 78.6       &  154                      & HD\,42807, HD\,46709      &\\
                    &             & $\geq0$    &          & SW            & 84.4       &  143                      & HD\,50692                 &\\
  VLTI/AMBER        & 2006-02-20  & 03:14      & K/LR     & UT1-UT3       & 101.0      &  40                       & HD\,45415                 &\\
  \noalign{\smallskip}
  \tableline
  \noalign{\smallskip}
  \multicolumn{9}{c}{\textbf{Mid-Infrared}}\\
  \noalign{\smallskip}
  \tableline
  \noalign{\smallskip}
  VLTI/MIDI         & 2004-10-30  & 08:49      & N/PRISM  & UT2-UT4       & 89.4       &  82                       & HD\,31421, HD\,49161      &\\
  VLTI/MIDI         & 2004-11-01  & 05:23      & N/PRISM  & UT2-UT4       & 55.9       &  90                       & HD\,25604, HD\,49161,     &\\
                    &             &            &          &               &            &                           & HD\,31421                 &\\
  VLTI/MIDI         & 2004-12-29  & 06:42      & N/PRISM  & UT2-UT3       & 46.5       &  46                       & HD\,49161                 &\\
  VLTI/MIDI         & 2004-12-30  & 02:33      & N/PRISM  & UT3-UT4       & 59.6       &  114                      & HD\,49161                 &\\
  VLTI/MIDI         & 2004-12-31  & 04:26      & N/PRISM  & UT3-UT4       & 61.5       &  108                      & HD\,49161                 &\\
  VLTI/MIDI (rej.)  & 2005-01-01  & 05:43      & N/PRISM  & UT3-UT4       & 54.7       &  106                      & HD\,31421, HD\,49161      &\\
  VLTI/MIDI (rej.)  & 2005-02-28  & 00:04      & N/PRISM  & UT2-UT3       & 39.3       &  44                       & HD\,49161                 &\\
  VLTI/MIDI         & 2007-02-08  & 05:41      & N/PRISM  & UT1-UT3       & 102.0      &  37                       & HD\,49161                 &\\
  VLTI/MIDI         & 2007-03-10  & 03:12      & N/PRISM  & UT1-UT2       & 56.4       &  35                       & HD\,49161                 &\\
\enddata
\tablerefs{
 (1) \citealt{mil01}; 
 (2) re-processing of data presented in \citealt{ake00}; 
 (3) \citealt{wil03}.
}
\end{deluxetable}


\begin{deluxetable}{lccccc}
\tabletypesize{\scriptsize}
\tablecaption{Calibrator star information for the interferometric observations presented in Tab.~\ref{tab:observations}.
  \label{tab:calibrators}}
\tablewidth{0pt}

\tablehead{
  \colhead{Star} & \colhead{$K$}  & \colhead{$N$}  & \colhead{Spectral}&  \colhead{$d_{\mathrm{UD},K}$}  & \colhead{$d_{\mathrm{UD},N}$}\\
  \colhead{}     & \colhead{}     & \colhead{[Jy]} & \colhead{Type}    &  \colhead{[mas]}                &  \colhead{[mas]}
}
\startdata
  \object{HD\,42807}  & 4.85 &  0.5  & G2V      & $0.45 \pm 0.03$\tablenotemark{a}&\\
  \object{HD\,43042}  & 4.13 &  1.0  & F6V      & $0.59 \pm 0.10$\tablenotemark{b}&\\   
  \object{HD\,43587}  & 4.21 &  1.0  & F9V      & $0.48 \pm 0.30$\tablenotemark{c}&\\
  \object{HD\,45415}  & 3.02 &  3.3  & G9III    & $1.06 \pm 0.02$\tablenotemark{d}&\\
  \object{HD\,46709}  & 2.62 &  4.4  & K5III    & $1.66 \pm 0.02$\tablenotemark{d}&\\
  \object{HD\,50692}  & 4.29 &  0.8  & G0V      & $0.56 \pm 0.10$\tablenotemark{b}&\\   
  \object{HD\,25604}  & 2.03 &  7.2  & K0III    &                       & $1.96 \pm 0.08$\tablenotemark{e}\\  
  \object{HD\,31421}  & 1.41 & 13.6  & K2IIIb   &                       & $2.58 \pm 0.15$\tablenotemark{e}\\  
  \object{HD\,49161}  & 1.58 &  7.2  & K4III    &                       & $2.88 \pm 0.17$\tablenotemark{e}\\  
\enddata
\tablecomments{The $V$-band magnitudes were taken from SIMBAD, the $K$-band magnitudes from the 2MASS point source catalog, and the $N$-band (12~$\mu$m) flux density from \citet{hel88}.}
\tablenotetext{a}{UD diameter from \citet{mal98}.}
\tablenotetext{b}{UD diameter from the CHARM catalog~\citep{ric02}.}
\tablenotetext{c}{UD diameter adopted from \citet{pas01}, using the Hipparcos parallax of 51.76~mas measured for HD\,43587.}
\tablenotetext{d}{UD diameter from \citet{mer05}.}
\tablenotetext{e}{UD diameter from the CHARM2 catalog~\citep{ric05}.}
\end{deluxetable}


\begin{figure}[tbp]
  \centering
  \includegraphics[angle=270,width=13cm]{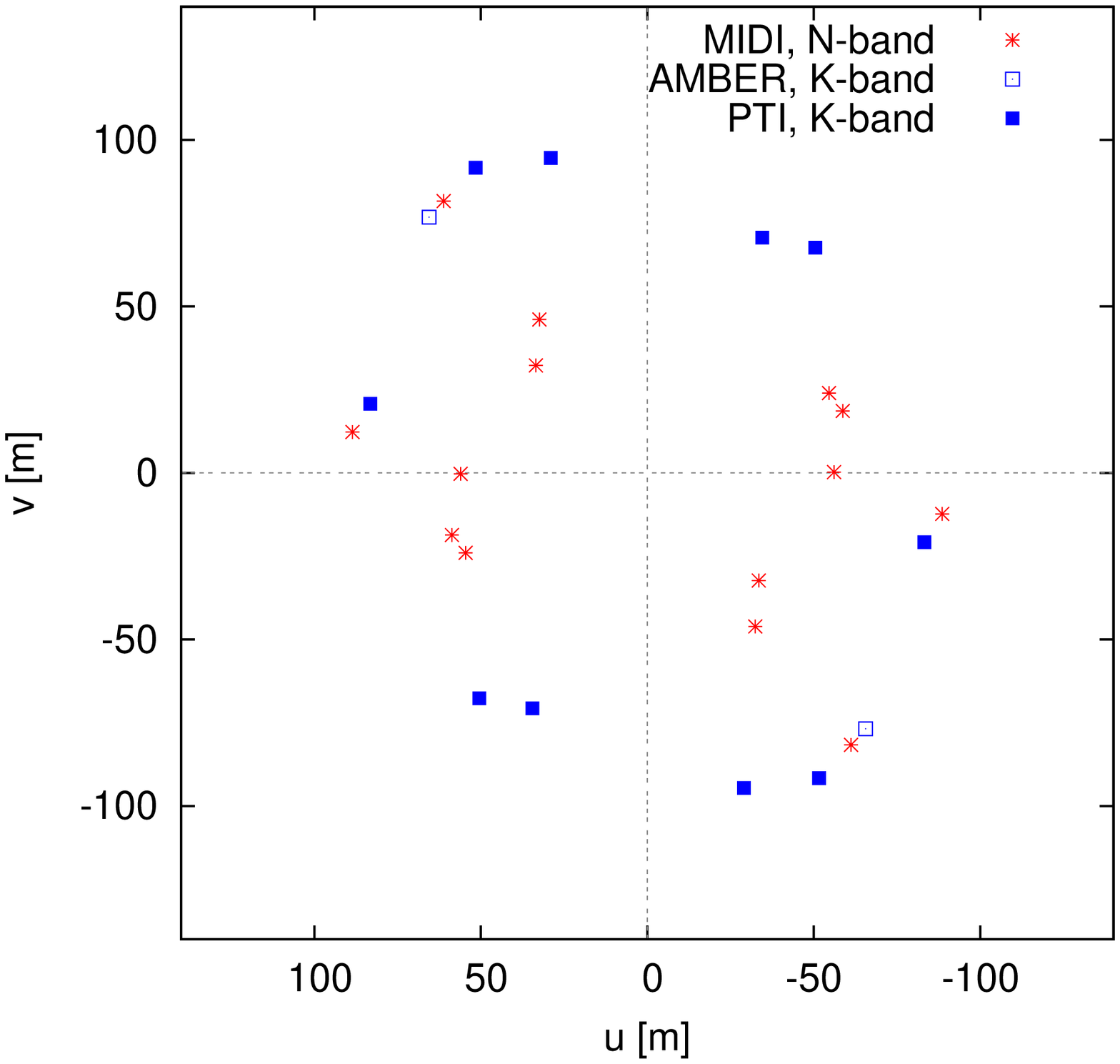}\\
  \caption{$uv$-plane coverage of the VLTI/MIDI, VLTI/AMBER, and archival PTI data.
    \label{fig:uvcov}
  }
\end{figure}


Details of all interferometric observations of MWC~147 used in this paper
are summarized in Tab.~\ref{tab:observations}.
All visibility measurements were
corrected for atmospheric and instrumental effects using calibrator stars
observed during the same night.
The calibrator stars as well as their assumed angular diameters are listed in
Tab.~\ref{tab:calibrators}.  Fig.~\ref{fig:uvcov} shows the $uv$-plane
coverage obtained with these observations.

\subsection{VLTI/MIDI observations}


\begin{figure}[tbp]
  \centering
  \includegraphics[angle=270,width=13cm]{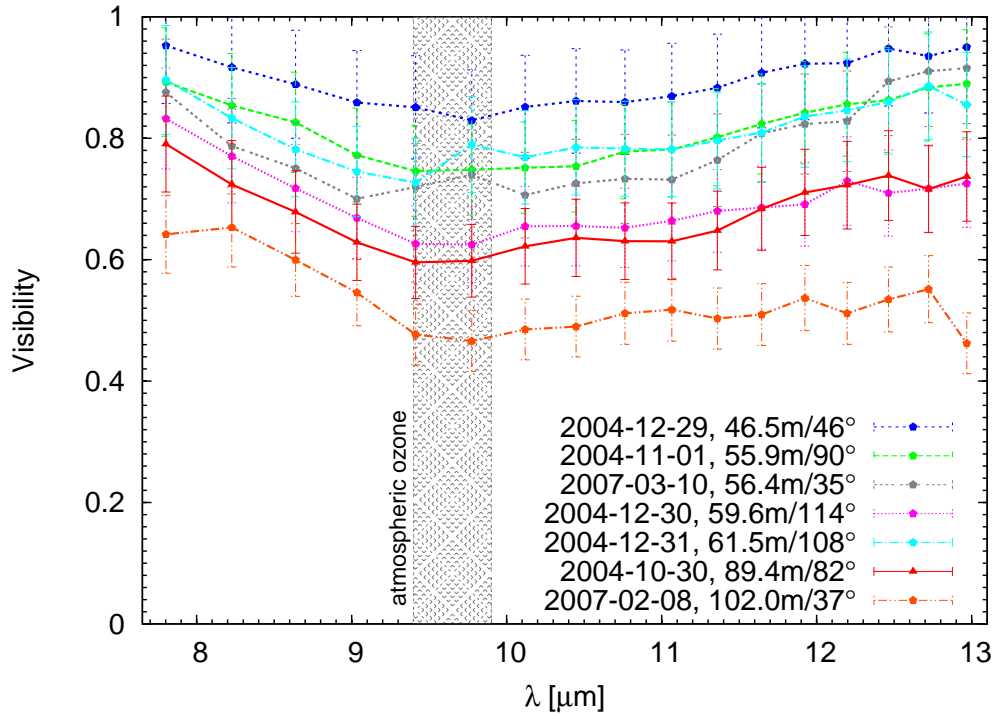}\\
  \caption{Visibilities measured with MIDI as a function of wavelength.
    \label{fig:VISMIDI}
  }
\end{figure}


The MIDI interferometer \citep{prz03, lei04}
at the ESO Very Large Telescope Interferometer (VLTI) records spectrally
dispersed interferograms in the $N$-band (8-13~$\mu$m).
The MIDI observations of MWC~147  were carried out 
for ESO open time (OT) programmes 074.C-0181 and 078.C-0129 
(P.I.\ Th.~Preibisch), using the NaCl prism as dispersive element (providing a 
spectral resolution of $R=30$) and the HIGH-SENS
instrument mode.  
In total, nine observations were carried out on five different baseline
configurations, as listed in Tab.~\ref{tab:observations}.

We used the data reduction
software package MIA+EWS\footnote{The MIA+EWS software package is available
  from the website \url{http://www.mpia-hd.mpg.de/MIDISOFT/}} (Release 1.5.1)
to extract visibilities from the MIDI data. 
This package contains two independent data reduction programs,
based on a coherent \citep[EWS, ][]{jaf04} and an incoherent approach
\citep[MIA, ][]{lei04}.
The reduction results obtained with both algorithms agree very
well (within 3\% for the calibrated visibility) for all data sets; 
with exception of the data sets from January/February 2005.
Inspection of the acquisition image for these specific data sets 
revealed a poor beam overlap; thus, we rejected them from our further analysis.
The inspection of the acquisition images for the other data sets
showed  that the visual companion at a separation of 3\farcs1~was not in the
MIDI field-of-view (FOV) and, therefore, does not affect the measured visibilities.
The wavelength-dependent calibrated visibilities of the remaining seven data sets
are shown in Fig.~\ref{fig:VISMIDI}.  
For the minimum relative error on the calibrated visibilities, we assume a conservative
value of 10\% \citep[see ][]{lei04}.

\subsection{VLTI/AMBER observations}


\begin{figure}[tbp]
  \centering
  \includegraphics[angle=270,width=13cm]{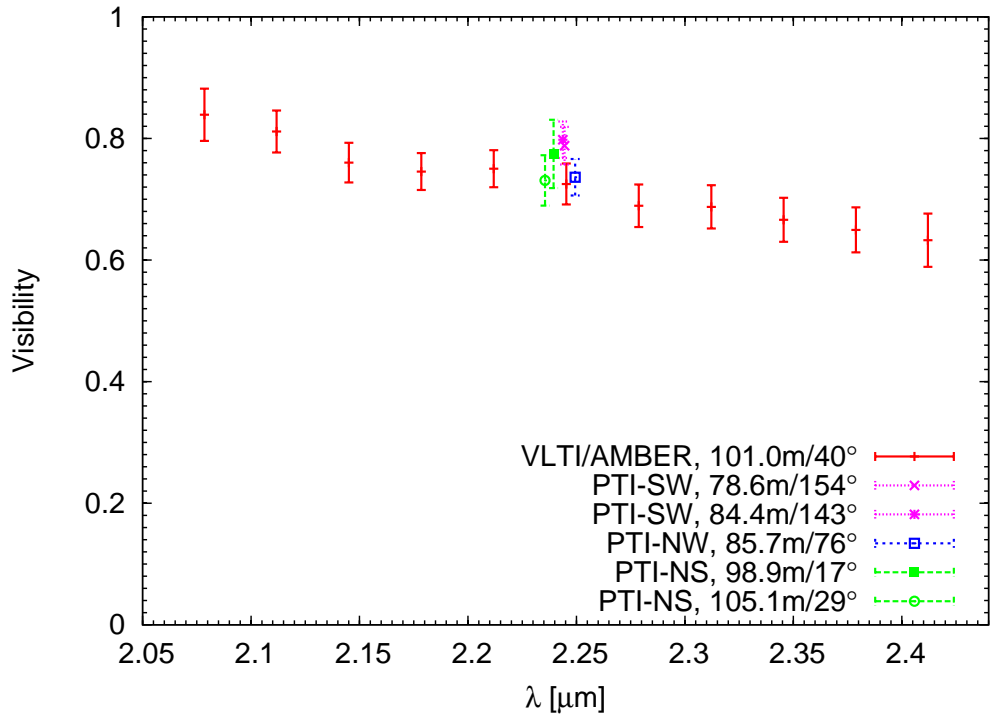}\\
  \caption{Wavelength-dependence of the visibility measured with VLTI/AMBER
    and the visibilities measured with PTI using broad-band filters.
    \label{fig:VISAMBER}
  }
\end{figure}


AMBER \citep{pet03, pet07}, the NIR beam-combiner of the VLTI, can combine the
light from up to three telescopes, providing not only the 
visibility amplitudes but also the closure phase relation. 
The observations of MWC~147 were conducted within OT programme 076.C-0138
(P.I.\ Th.~Preibisch) using the 8.2~m unit telescopes UT1-UT3-UT4. 
Spectrally dispersed interferograms were obtained 
in the low resolution (LR) mode ($R=35$), 
which resolves the $K$-band into 11 spectral channels.
Due to
problems with the fiber injection during that night, the flux reaching the
AMBER beam combiner from UT4 was about a factor of 3 lower than from the other
telescopes.  Therefore, clear fringes were detectable on only one of the three
baselines (UT1-UT3), and no closure phase signal could be measured.  
Following the ESO service mode observation procedure, one data set of
5000~interferograms was recorded both on a calibrator star (HD\,45415) and
on MWC~147.
The length and orientation of the projected baseline for this AMBER measurement
(101~m, PA 40$^{\circ}$) is similar to the measurement at the PTI-NS baseline,
but adds information about the spectral dependence of the visibility within
the $K$-band. 

The AMBER data were reduced with version 2.4 of the
\textit{amdlib}\footnote{The amdlib software package is available from the
  website \url{http://amber.obs.ujf-grenoble.fr}} software employing the P2VM
algorithm~\citep{tat06}. 
Due to the absence of a fringe tracker, a large fraction of the interferograms
is of rather low contrast \citet[see discussion in][]{pet07}.  Therefore, we
removed those frames from our data set for which \textit{(a)} the light
injection from the contributing telescopes was unsatisfying; i.e., the
intensity ratio between the photometric channels was larger than 4, 
\textit{(b)} the atmospheric piston was larger than $^1\!/\!_4$ of the coherence
length $\lambda \cdot R$, or
\textit{(c)} the fringe contrast was decreased due to instrumental jitter 
(the 20\% best interferograms were selected based on the Fringe SNR criteria,
as defined in \citealt{tat06}).
In Fig.~\ref{fig:VISAMBER} the calibrated $K$-band visibilities
derived from the AMBER and PTI measurements are shown as a function of wavelength.
As mentioned in \citet{pet07}, the accuracy of the absolute
calibration of the visibilities measured with VLTI/AMBER is currently
limited by vibrations induced by the UTs, which lead us to assume a
minimum relative error of 3\% for the calibrated visibilities.
In contrast to the absolute calibration, the wavelength-differential
dependence of the visibility is insensitive to this effect \citep{pet07}.

\subsection{PTI archive data}

MWC~147 was observed with the Palomar Testbed Interferometer \citep[PTI,~][]{col99}
on the NS \citep{ake00} and NS \& NW baselines \citep{wil03}.  
Yet unpublished data for the SW baseline was retrieved from the PTI archive.
To obtain a uniformly calibrated data set, we processed the new data set
together with the previously published data using the V2Calib
V1.4 software\footnote{The V2Calib software is available from the website
  \url{http://msc.caltech.edu/software/V2calib/}}.
In the course of the calibration procedure, we applied to the raw visibilities
(which were estimated from the spatially filtered PTI spectrometer output)
the standard correction for coherence loss using the measured phase jitter
\citep[see][]{col99}.
The individual PTI measurements were binned so that each bin contains 
data sets covering less than $15^\circ$ variation along the PA.
Since the measurements on the PTI-NS baseline also cover a relatively wide
range of PAs ($\sim 23^\circ$), we divided those measurements into two halves
(depending on the PA) before averaging.
As for the AMBER data, we assume 3\% minimum relative error on the
calibrated visibilities.

\subsection{Spitzer-IRS archive data}

In order to constrain the SED for our radiative transfer modeling as
tightly as possible, we obtained MIR spectra from the \textit{Spitzer} Space
Telescope Archive.  These spectra were recorded on 2004-10-26 within GTO
programme ID~3470 (P.I.\ J.~Bouwman) using the Infrared Spectrograph
\citep[IRS,][]{hou04}.
The data set consists of four exposures; two taken in the Short-High mode
(SH, wavelength range from 9.9 to 19.6~$\mu$m) and two in the Long-High
mode (LH, 18.7 to 37.2~$\mu$m).  Both modes provide a spectral resolution
of $R\sim 600$.
With slit sizes of $4\farcs7 \times 11\farcs3$ (SH mode) and $11\farcs1 \times
22\farcs3$ (LH mode), IRS integrates flux from areas much
larger than those collected in the spatially filtered
MIDI spectrum.
The spectra were pre-processed by the S13.2.0 pipeline version at the Spitzer
Science Center (SSC) and then extracted with the SMART software, Version
5.5.7 \citep{hig04}.

\section{Results}

\subsection{The MIR spectrum}

\clearpage

\begin{figure}[tbp]
  \centering
  \includegraphics[angle=270,width=13cm]{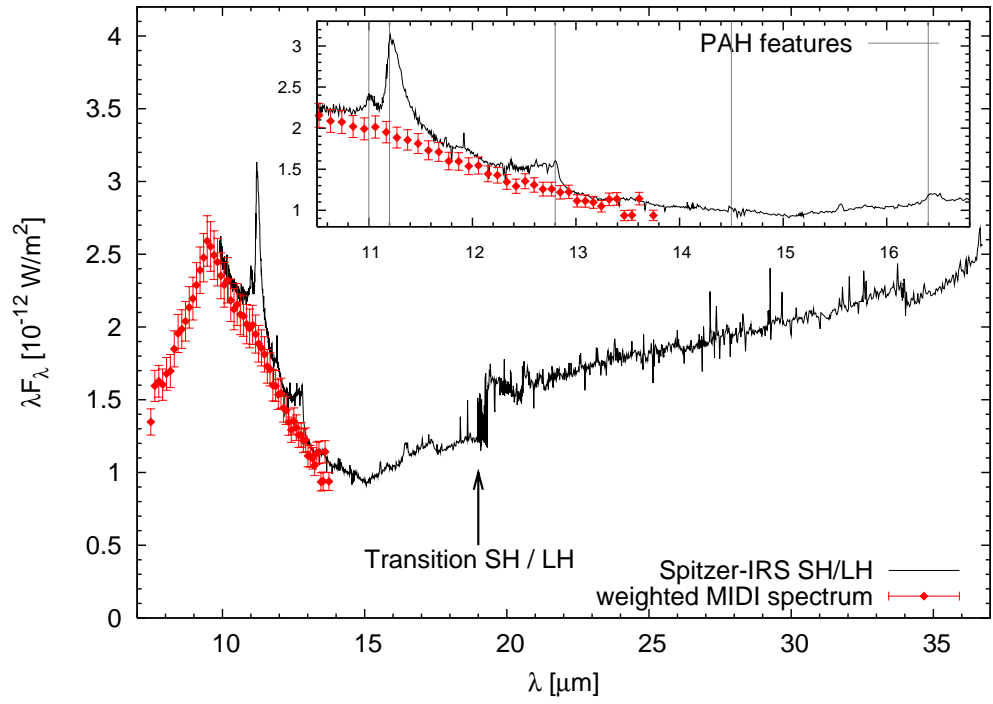}\\
  \caption{Comparison of the measured MIDI spectrum and the spectrum extracted
    from archival \textit{Spitzer}-IRS data.
    \label{fig:MIDIIRSspectrum}
  }
\end{figure}

\clearpage

In the overlapping wavelength regime between 10 and 13~$\mu$m,
the MIDI and \textit{Spitzer}-IRS spectra show good quantitative
agreement, both in the absolute level of the continuum flux and in the
spectral slope (see Fig.~\ref{fig:MIDIIRSspectrum}). 
However, the IRS spectrum exhibits some line features which do not appear in
the MIDI spectrum\footnote{In order to validate that the non-detection of the
strong emission line around 11.2~$\mu$m in the MIDI spectrum is not only due
to MIDI's lower spectral resolution, we convolved the
\textit{Spitzer}-IRS spectrum to the resolution of the MIDI PRISM.  Since the 
line is still clearly visible in the convolved spectrum, we are confident
that no significant line emission is present in the spatially filtered
MIDI spectrum.}.
As these emission lines are most pronounced at wavelengths of 11.0, 11.2,
12.8, 14.5, and 16.4~$\mu$m, we attribute these features to the presence of
Polycyclic Aromatic Hydrocarbons \citep[PAHs, ][]{all85, dis04},
which were found towards a large variety of objects, 
including T-Tauri stars and HAeBe stars \citep{ack04}.
For the strong and rather broad emission feature at 11.2~$\mu$m, contributions
from the 11.3~$\mu$m crystalline silicate feature are also possible.

The prominence of the PAH lines in the {\it Spitzer}-IRS spectra with their
$4\farcs7 \times 11\farcs3$ FOV 
and their
absence in the MIDI spectrum with its much ($\sim 17\times$)
smaller FOV suggests that the PAH emission comes predominantly from the
outermost circumstellar environment and/or the surrounding
nebulosity of MWC~147, similar to what was found for other
young stellar objects (e.g.\ \citealt{van04b}, \citealt{rho06}) or the outer
regions of HAeBe disks \citep{hab04, hab06}.

The comparison of the {\it Spitzer}-IRS spectrum with the fluxes
measured by IRAS \citep{hel88} showed that the
IRAS fluxes are systematically higher.  This is likely related to the
larger beam size of IRAS, which includes
significant amounts of emission from the ambient NGC\,2247 nebula.
For our modeling in Sect.~\ref{cha:RT}, we will therefore 
treat the IRAS fluxes as upper limits only.

\subsection{The correlated MIR spectrum -- indications of grain growth}\label{cha:intgraingrowth}


\begin{figure}[tbp]
  \centering
  \includegraphics[angle=270,width=13cm]{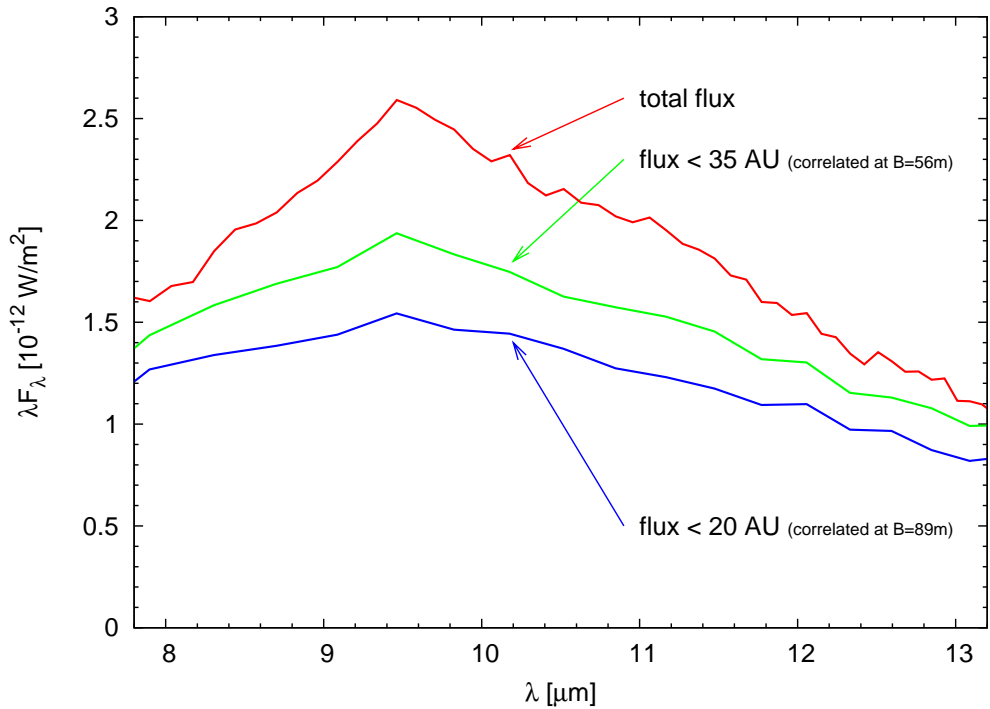}\\
  \caption{Total MIDI spectrum and the correlated flux at the 55.9~m
    (corresponding to a spatial scale of $\lesssim 35$~AU) and 89.4~m
    ($\lesssim 20$~AU) baselines.
    \label{fig:corrflux}
  }
\end{figure}


The visibilities measured with MIDI show significant variations along the
recorded wavelength range. In particular, we detect a drop of visibility
within the $10~\mu$m silicate feature.
A similar behavior has already been observed for several HAeBe stars; e.g.,
in the samples of \citet{lei04} and \citet{van04a}. 

As the silicate emission feature is generally attributed to the presence of
rather small silicate grains \citep[$r \lesssim 0.1~\mu$m, see e.g.~][]{van04a},
it is possible to probe the radial dust mineralogy by comparing the
correlated spectrum at various baseline lengths with the total spectrum
$F_{\mathrm{tot}}$.  The correlated spectrum $F_{\mathrm{corr}}$
corresponds to the flux integrated over the spatial area unresolved
by the interferometer for a particular baseline length $B$. Therefore, for
each baseline length $B$, the correlated flux $F_{\mathrm{corr}}(B)$ can be
computed by 
multiplying the total spectrum measured by MIDI in the photometry files with
the visibility measured for a certain baseline.  In order to probe the radial
dependence of the dust mineralogy, measurements taken at similar PAs should be
used in order to avoid contaminations by changes in the source geometry.  We
therefore choose the measurements from 2004-11-01 ($B=55.9$~m, PA=82\degr) and
2004-10-30 ($B=89.4$~m, PA=90\degr), as they have very similar PAs. The
comparison of the correlated spectra for these baselines with the total 
spectrum (Fig.~\ref{fig:corrflux}) shows that the 10~$\mu$m silicate feature
flattens out with increasing resolution. This change in the correlated
spectrum might indicate spatial variations in the dust composition, with the
more evolved dust grains \citep[i.e.\ larger grains with a weaker silicate 
feature;~][]{min06} in the innermost disk regions.

\subsection{Geometric model fits} \label{cha:resgeomodelfit}

Since the imaging capabilities of the current generation of infrared
interferometers are rather limited, 
the measured interferometric observables are often used 
to constrain the parameters of a model for the
object morphology.
In most studies presented until now, either purely geometric profiles (in
particular uniform disk (UD) and Gaussian profiles) or physically motivated
geometries such as ring profiles or analytic accretion disk models with a
power law temperature distribution were employed.  Ring models are justified by the
theoretical expectation that most of the NIR emission originates from a rather
small region around the dust sublimation radius 
\citep[e.g.,\ ][]{mil01, mon02}.

A common problem in applying simple geometric models is that the observed
emission does not originate exclusively from the circumstellar material:  a
certain fraction comes directly from the central star and contributes a
spatially (nearly) unresolved component, and the existence of extended
background emission, which is fully resolved, is also possible.
For the model fits, one therefore has to specify which fraction of the total
flux $F_\mathrm{tot}$ at any wavelength has to be attributed to the different
spatial components.  The stellar flux contribution 
$f_\mathrm{star/tot}(\lambda)=F_\mathrm{star}/F_\mathrm{tot}$ is often
estimated from the SED, while the extended component
$f_\mathrm{ext/tot}(\lambda)=F_\mathrm{ext}/F_\mathrm{tot}$ is usually
assumed to be  zero. These assumptions are, however, 
associated with a considerable uncertainty.

\begin{figure}[tbp]
  \centering
  \includegraphics[angle=270,width=13cm]{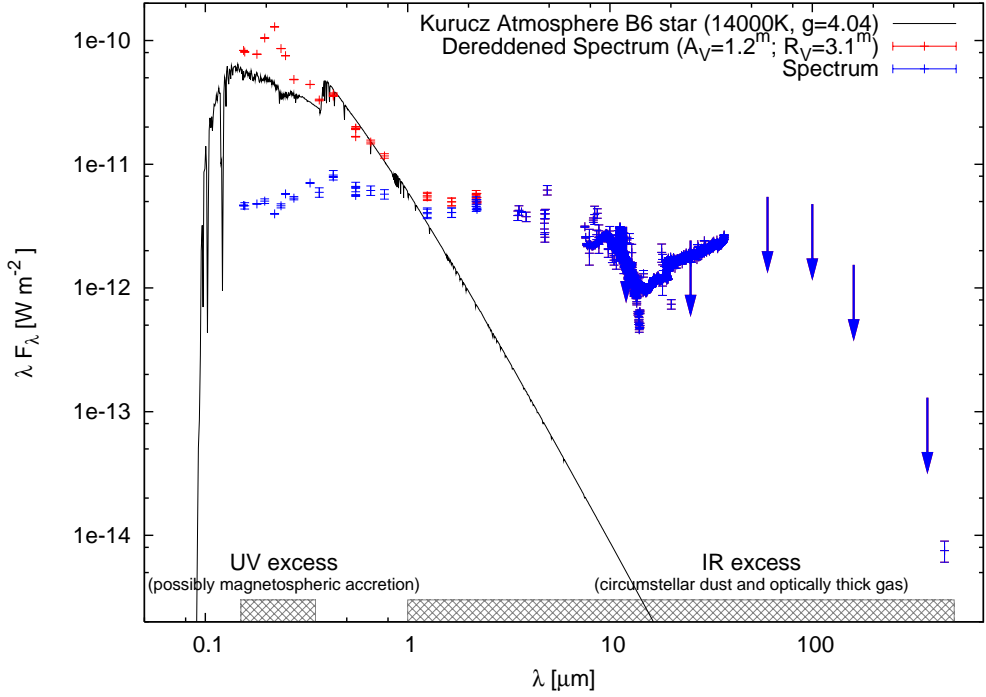}\\
  \caption{
    SED for MWC\,147 containing the measured averaged, weighted MIDI
    spectrum, the archival \textit{Spitzer}-IRS spectrum and data from
    literature.  Comparing the de-reddened SED with the Kurucz stellar
    atmosphere model for a B6-star reveals excess emission both in the UV and
    IR wavelength regime.  The UV excess emission might originate in magnetospheric
    accretion (see the general models by \citealt{muz04}), while the infrared excess
    likely indicates the presence of circumstellar dust and optically thick
    gas (see modeling in Sect.~\ref{cha:RT}).
    We included photometric data from
    \citet[][ 156.5~nm, 196.5~nm, 236.5~nm, 274.0~nm]{tho78},
    \citet[][ 150~nm, 180~nm, 220~nm, 250~nm, 330~nm]{wes82},
    \citet[][ $U, B, V, RC, IC, J, H, K, L, M, N, Q$-bands]{hil92},
    \citet[][ $V$-band]{tur93}, 
    \citet[][ $B$-band]{egr92},
    \citet[][ $V_T$, $B_T$, $R_J$, $I_J$-band]{hog00},
    \citet[][ 2MASS, $J$, $H$, $K_s$-band]{skr06},
    \citet[][ $2.2~\mu$m, $3.5~\mu$m, $3.65~\mu$m, $4.8~\mu$m, $4.9~\mu$m, $8.4~\mu$m, $8.6~\mu$m, $10.8~\mu$m, $11.0~\mu$m, $11.3~\mu$m, $12.8~\mu$m, $18~\mu$m]{coh73a},
    \citet[][ $4.74~\mu$m, $7.91~\mu$m, $8.81~\mu$m, $10.27~\mu$m, $11.70~\mu$m, $12.49~\mu$m, $N$-band, $18.17~\mu$m]{pol02},
    \citet[][ MSX 2.6, $4.29~\mu$m, $8.28~\mu$m, $12.13~\mu$m, $14.65~\mu$m, $21.34~\mu$m]{ega99},
    \citet[][ $2.2~\mu$m, $3.85~\mu$m, $8.65~\mu$m, $9.97~\mu$m, $10.99~\mu$m, $11.55~\mu$m]{ber87},
    \citet[][ $12~\mu$m, $25~\mu$m, $60~\mu$m, $100~\mu$m]{hel88},
    \citet[][ $160~\mu$m, $370~\mu$m]{cas91},
    \citet[][ $450~\mu$m]{man94}.
    \label{fig:SED}
  }
\end{figure}

To allow comparison with earlier NIR interferometric studies on MWC~147,
we keep the flux ratios from \citet{mil01}, namely
$f_\mathrm{star/tot}(2.1~\mu$m$)=0.16$, and $f_\mathrm{ext/tot}(2.1~\mu$m$)=0.0$ for the
analytic fits. The same flux ratio was assumed by \citet{wil03}, while
\citet{ake00} used $f_\mathrm{star/tot}(2.1~\mu$m$)=0.10$.  
At MIR wavelengths, the stellar contribution is likely to be 
negligible; i.e., $f_\mathrm{star/tot}(10~\mu$m$)\approx 0$.  This can be concluded from
the SED shown in Fig.~\ref{fig:SED}, where the infrared excess exceeds the
stellar flux by a factor of $\sim 280$ at $10~\mu$m.

\subsubsection{Characteristic source size and elongation}\label{cha:resgeomodelfitelong}


\begin{deluxetable}{lccccccccccc}
\tabletypesize{\scriptsize}
\tablecaption{Model fits for circular geometries.
\label{tab:sizes}}
\tablewidth{0pt}

\tablehead{
  \colhead{}   & \multicolumn{2}{c}{$K$-band} &\colhead{}& \multicolumn{2}{c}{$9~\mu$m} &\colhead{}& \multicolumn{2}{c}{$11~\mu$m} &\colhead{}& \multicolumn{2}{c}{$12.5~\mu$m}\\
\cline{2-3} \cline{5-6} \cline{8-9} \cline{11-12} \\
  \colhead{}   &  \colhead{Diameter}          & \colhead{$\chi^2_r$} &\colhead{}& \colhead{Diameter} & \colhead{$\chi^2_r$} &\colhead{}& \colhead{Diameter}        & \colhead{$\chi^2_r$}   &\colhead{}& \colhead{Diameter}       & \colhead{$\chi^2_r$}\\
  \colhead{}   &  \colhead{[mas]}             & \colhead{}           &\colhead{}& \colhead{[mas]}    & \colhead{}           &\colhead{}& \colhead{[mas]}           &  \colhead{}            &\colhead{}& \colhead{[mas]}          & 
}
\startdata
  Uniform Disk (UD)        & $2.6\pm0.2$ & 1.44         && $12.8\pm2.0$ & 1.38        && $17.3\pm2.2$ & 0.95        && $18.1\pm2.2$ & 0.89\\
  Gaussian                 & $1.6\pm0.1$ & 1.30         && $7.9\pm1.4$  & 1.19        && $10.8\pm1.5$ & 0.75        && $11.2\pm1.5$ & 0.90\\
  Ring                     & $1.8\pm0.1$ & 1.51         && $8.7\pm1.4$  & 1.49        && $11.8\pm0.3$ & 1.07        && $12.4\pm1.3$ & 0.90\\
\enddata
\tablecomments{For the $K$-band fits, we attribute 16\% of the total flux to the unresolved stellar component.  At a distance of 800~pc, 1~mas corresponds to 0.8~AU.}
\end{deluxetable}

\begin{deluxetable}{lccccccccccccccccccc} \rotate
\tabletypesize{\scriptsize}
\tablecaption{
  Model fits for inclined geometries.
\label{tab:sizesincl}}
\tablewidth{0pt}

\tablehead{
  \colhead{}   & \multicolumn{4}{c}{$K$-band} &\colhead{}& \multicolumn{4}{c}{$9~\mu$m} &\colhead{}& \multicolumn{4}{c}{$11~\mu$m} &\colhead{}& \multicolumn{4}{c}{$12.5~\mu$m}\\
\cline{2-5} \cline{7-10} \cline{12-15} \cline{17-20} \\
  \colhead{}   &  \colhead{Diameter} & \colhead{$i$}      & \colhead{PA}         & \colhead{$\chi^2_r$} &\colhead{}& \colhead{Diameter} & \colhead{$i$}      & \colhead{PA}         & \colhead{$\chi^2_r$} &\colhead{}& \colhead{Diameter}        & \colhead{$i$}      & \colhead{PA}         & \colhead{$\chi^2_r$}   &\colhead{}& \colhead{Diameter}       & \colhead{$i$}      & \colhead{PA}        & \colhead{$\chi^2_r$}\\
  \colhead{}   &  \colhead{[mas]}    & \colhead{[$^\circ$]} & \colhead{[$^\circ$]} & \colhead{}           &\colhead{}& \colhead{[mas]}    & \colhead{[$^\circ$]} & \colhead{[$^\circ$]} & \colhead{}           &\colhead{}& \colhead{[mas]}           & \colhead{[$^\circ$]} & \colhead{[$^\circ$]} &  \colhead{}            &\colhead{}& \colhead{[mas]}          & \colhead{[$^\circ$]} & \colhead{[$^\circ$]}& 
}
\startdata
UD                & $3.0 \times 2.4$ & 39 & 12 & 0.33  && $20.1 \times 10.8$ & 58 & 58 & 1.04 && $20.8 \times 15.3$ & 43 & 53 & 0.59 && $23.5 \times 15.7$ & 48 & 70 & 0.58\\
Gaussian          & $1.8 \times 1.4$ & 39 & 11 & 0.30  && $18.5 \times 6.5$ &  69 & 66 & 0.89 && $23.3 \times 8.9$  & 68 & 65 & 0.31 && $23.5 \times 9.0$  & 68 & 74 & 0.49\\
Ring              & $2.1 \times 1.6$ & 41 & 12 & 0.34  && $14.0 \times 7.5$ &  58 & 60 & 1.06 && $20.2 \times 10.1$ & 60 & 66 & 0.48 && $16.3 \times 10.9$ & 48 & 64 & 0.59\\
\enddata
\tablecomments{
  The angle $i$ denotes the inclination computed from the fitted axis
  ratio, assuming an underlying circular source structure.  However, due to
  the unknown flux contribution from the unresolved central region (i.e.\ 
  stellar flux, for which we assume $f_\mathrm{star/tot}(\mathrm{NIR})=0.16$
  and $f_\mathrm{star/tot}(\mathrm{MIR})=0.0$, and maybe additional accretion
  luminosity), and the incomplete PA-coverage of the MIDI data set, we
  consider this value rather uncertain.
}
\end{deluxetable}


\begin{figure}[tbp]
  \centering
  \begin{minipage}{12cm}
    \centering
    $\begin{array}{cc}
      \hspace{-20mm}\includegraphics[width=82mm]{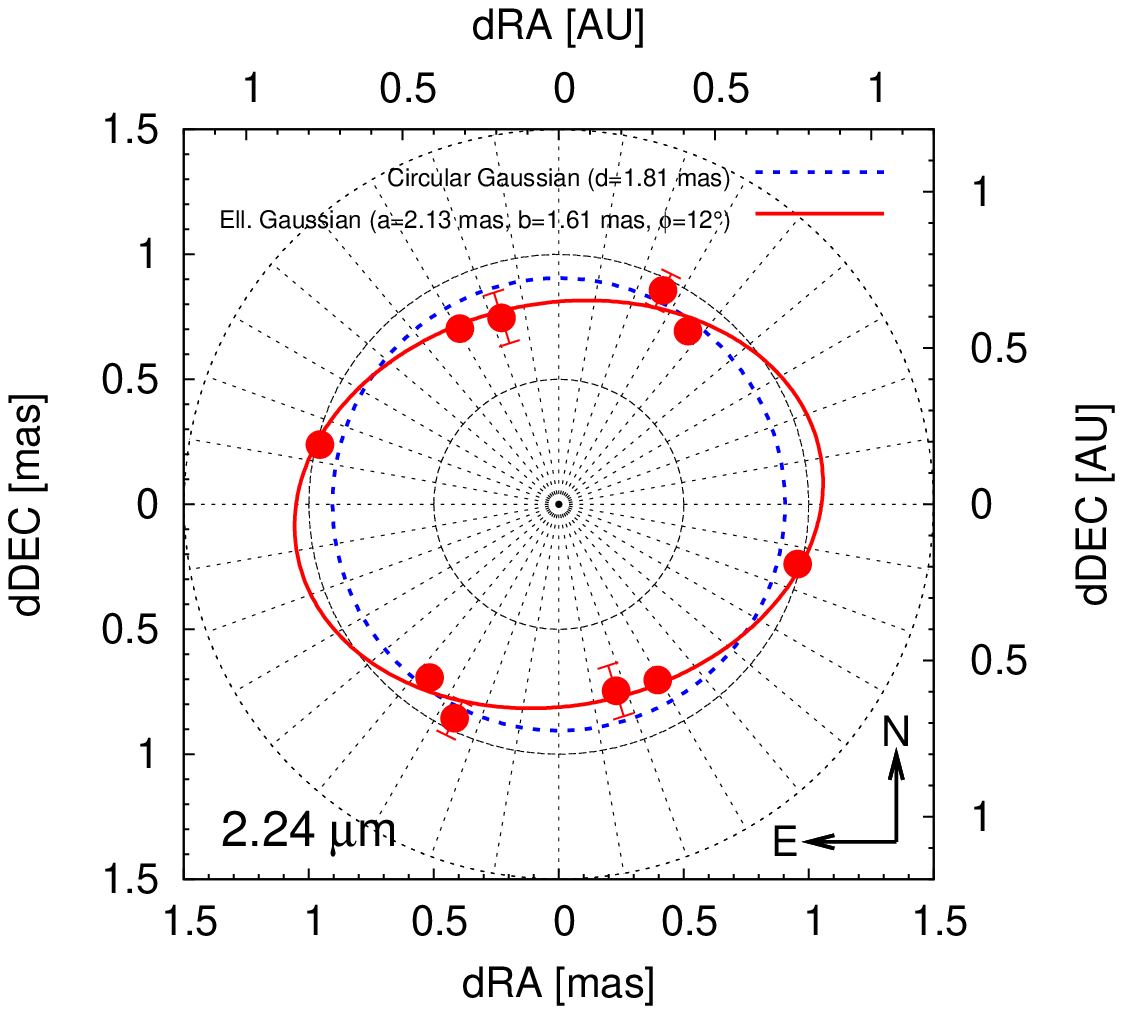} & \includegraphics[width=78mm]{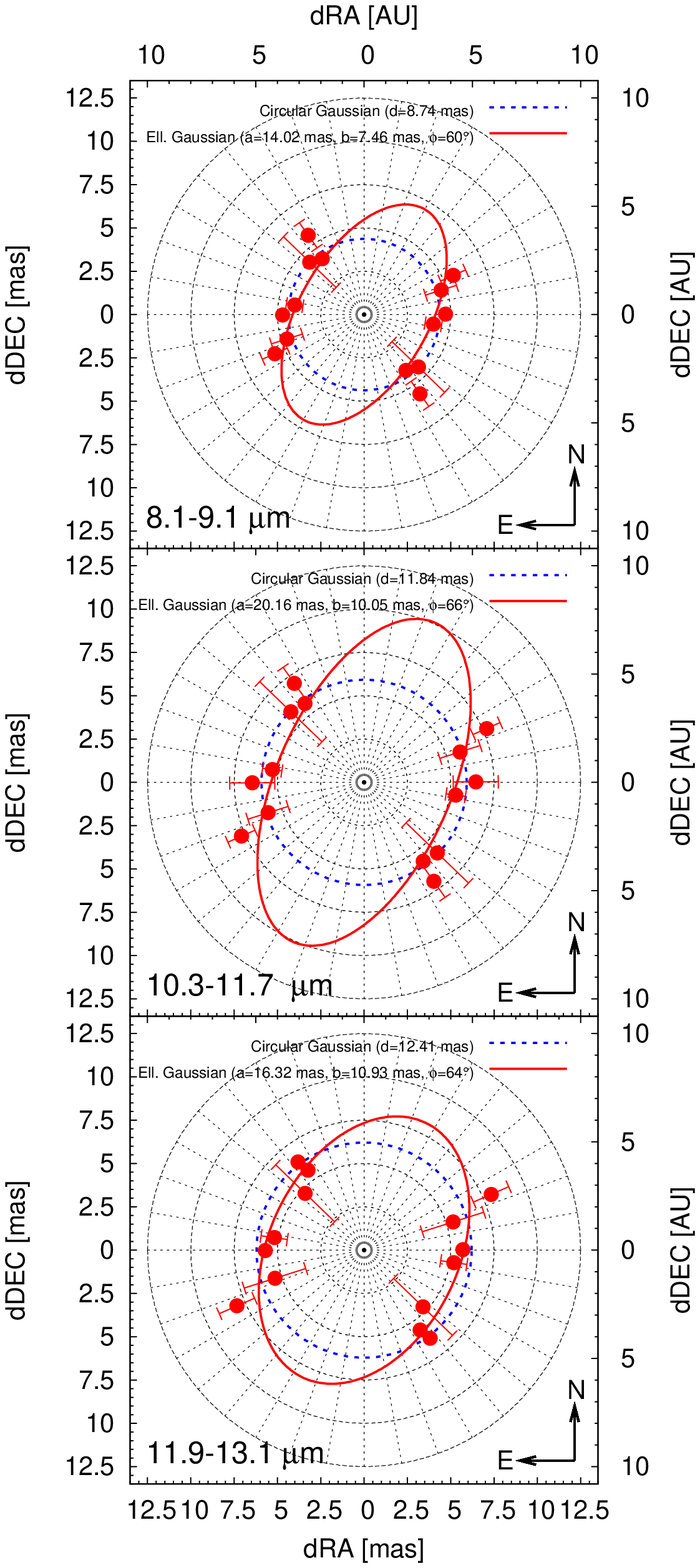}
    \end{array}$
  \end{minipage}

  \caption{Polar diagram showing the best-fit circular and elliptical
    geometries.  To derive the characteristic object size along different PAs
    (dots with errorbars), we fitted ring profiles to the PTI $K$-band
    visibilities (left panel) and to MIDI visibilities (right panel) which
    were averaged over five spectral channels around the silicate feature
    ($10.3-11.7~\mu$m) and in the surrounding continuum ranges $8.1-9.1~\mu$m
    and $11.9-13.1~\mu$m.
    \label{fig:fitellipse}
  }
\end{figure}


\begin{figure}[tbp]
  \centering
  \includegraphics[angle=270,width=13cm]{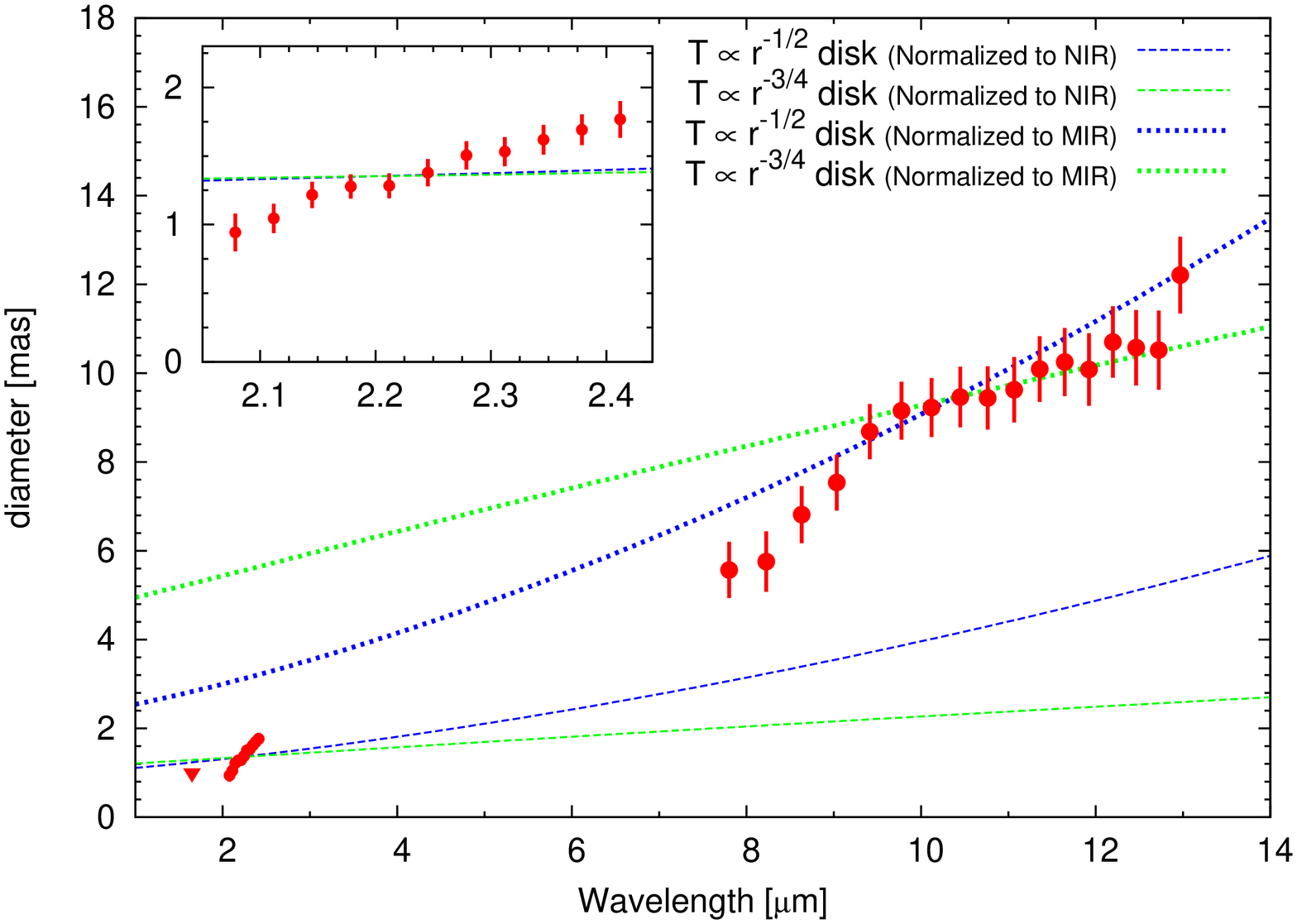}\\
  \caption{
    Wavelength-dependence of the measured characteristic size over the
    \textit{H}-, \textit{K}- and \textit{N}-band, including IOTA, PTI, AMBER
    and MIDI measurements.  From our seven MIDI measurements, we choose the data
    set from 2007-02-08, since this measurement was taken at
    very similar baseline length and PA (102.0~m/35\degr) as the AMBER data
    set (101.0~m/40\degr), which makes these data sets
    particularly well suited to study the radial disk structure without
    contamination by the detailed source geometry.
    The IOTA $H$-band measurement provides only an upper size limit.  
    To compute the characteristic size, Gaussian intensity profiles were
    assumed. 
    For comparison, we show the wavelength-dependent size corresponding to the
    commonly applied analytic disk model from \citet{hil92}.
    We scaled these disk models to match the measured NIR (dashed
    curves) or MIR size (dotted curves).  In all cases, it is evident that
    these analytic models cannot describe the measured wavelength-dependent
    size well.
    \label{fig:diamvariation}
  }
\end{figure}


To obtain a first estimate of the object size, we fit the most common
analytic profiles to our interferometric data: Gaussian, UD, and
ring profiles. For mathematical descriptions of these profiles, we refer to
\citet[][ UD profile]{kra05} and \citet[][ Gaussian \& ring profile]{mil01}.
We assume uniform bright rings with an average diameter $\Theta$ and a fixed 
width of 20\% to be consistent  with \citet{mon05}.
As the apparent object size is expected to change with wavelength, we
fitted these profiles to subsets of our data, covering wavelength ranges
around $2.2~\mu$m, $8.6~\mu$m, $11.0~\mu$m, and $12.5~\mu$m.
The visibilities measured in these sub-bands were fitted to the visibility
profiles using a Levenberg-Marquardt least-square fitting algorithm, taking
the chromatic change in resolution within the bandwidth into account.

The fits were performed for the case of circular symmetry (e.g.\ a disk seen
face-on) and for elliptical structures (e.g.\ an inclined disk). The
obtained diameters and goodness-of-fit values 
($\chi^2_r = \sum \left[ (V^2-V^2_{m})/\sigma_V \right]^2/N_{V}$, with $V^2$
being the measured squared visibility, $V^2_{m}$ the squared visibility computed
from the model, and $N_{V}$ the number of measurements) are given
in Tab.~\ref{tab:sizes} and~\ref{tab:sizesincl}.\\
The $\chi^2_r$ values already indicate that the elliptical geometries are a
better representation of our data than the circular models.
In order to illustrate this object elongation, we show the corresponding
geometries in Fig.~\ref{fig:fitellipse}.  Since the detection of object
elongation requires strict uniformity in the observational methodology, 
we did not include the single-baseline AMBER measurement because the mixture
of broadband and spectrally dispersed interferometric observations might
easily introduce artefacts. 

While the elongation is only marginally evident in the PTI NIR measurements, 
it is more significant in the MIR data, although the limited
PA-coverage of the available MIDI data (covering $\sim 80$\degr~in PA)
prevents us from measuring the precise axis ratio.
The effect that the deviations from circular geometry seem to be stronger at MIR
wavelengths might indicate that for the NIR model fits, the assumed stellar
flux ($f_\mathrm{star/tot}(2.1~\mu$m$)=0.16$) underestimates the real flux
contributions from a spatially unresolved region (in agreement with our
radiative transfer modeling results in Sect.~\ref{cha:RTDISKACC}).  
For the MIR-interferometric data, it is interesting to note that the
elongation found agrees well with the one seen in the $11.7/18.2~\mu$m color
temperature map published by \citet{pol02}.  Although these color temperature
maps show structures on scales of several arcseconds (e.g., on scales a
hundred times larger than our interferometric data), their orientation ($\sim
50^\circ$) and rough axis ratio are similar to those of the structure seen in
our MIDI observations.

\subsubsection{Wavelength-dependent characteristic size}

Our combined MIDI/AMBER/PTI/IOTA data set also allows us to study the
wavelength-dependence of the apparent size. For this, we fitted the
visibility measurement in each individual spectral channel with the analytic
formula for Gaussian intensity profiles (the result does not depend
strongly on the assumed profile).
The determined diameters are shown in Fig.~\ref{fig:diamvariation}.
The increase of the apparent size with wavelength is usually interpreted as a
consequence of the radial temperature profile for the circumstellar material
(i.e.~material at larger distances from the star is cooler).
However, as will be qualitatively discussed in Sect.~\ref{cha:intsizewlen},
the increase of the characteristic size with wavelength is much steeper than
expected for single power-law temperature accretion disks with an inner
hole, suggesting that these models cannot successfully explain the observed
wavelength-dependence of the emitting structure. 
In the context of our radiative transfer modeling in Sect.~\ref{cha:RT}, we
will discuss that this discrepancy can be explained assuming the presence of
an additional, more compact emitting region which dominates the emission at
NIR wavelengths, but whose contribution becomes less important at
MIR wavelengths. 

\citet{ake00} examined the possibility whether the measured visibilities
could indicate the presence of a close companion.  Since PTI observed
MWC~147 several times over a period of $\sim4$~yrs and found no significant
variations on the NS-baseline, we consider the binary scenario as very
unlikely.  For example, assuming, just for the sake of argument, a total
system mass of 7~$M_{\sun}$ and a semi-major axis of 10~AU, this would give an
orbital period of $\sim 12$~yrs and should result in significant visibility
variations over the covered 4~yrs.

\subsection{Comparison with analytic disk models}\label{cha:intsizewlen}

\begin{figure}[tbp]
  \centering
  \includegraphics[angle=0,width=17cm]{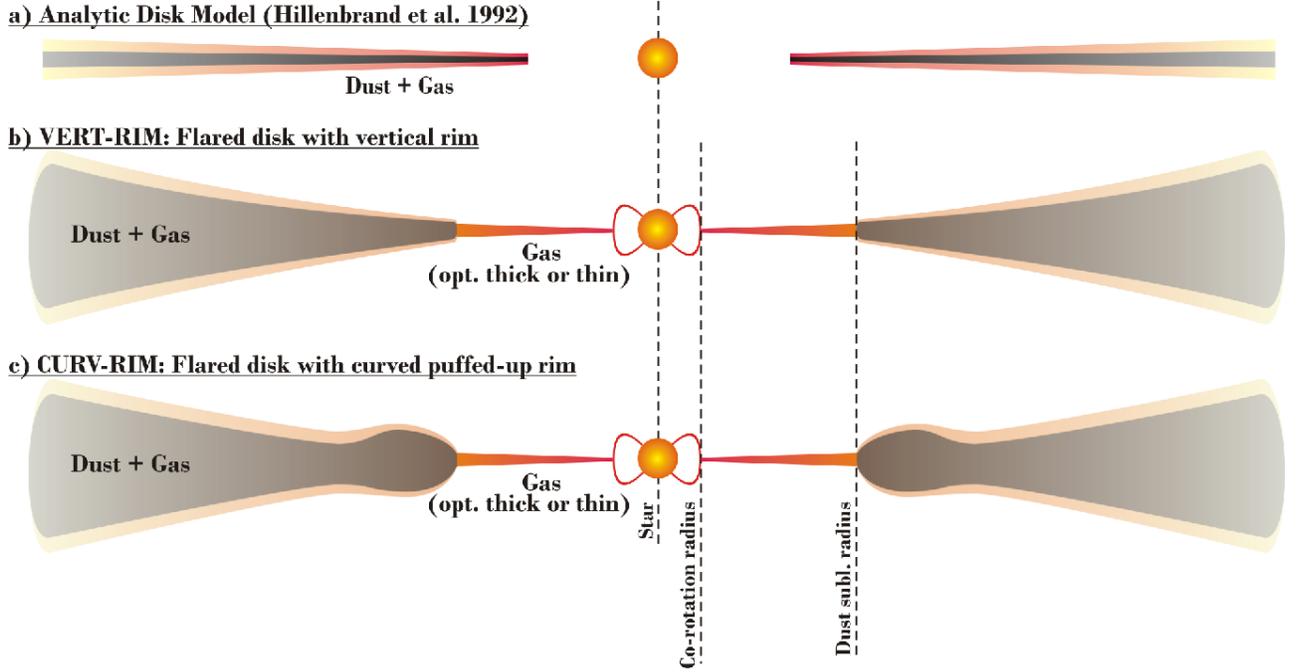}\\
  \caption{
    Illustration of the considered disk model geometries: 
    {\it(a)} ``Classical'' accretion disk geometry as considered in the 
    \citet{hil92} model (see Sect.~\ref{cha:intsizewlen}), incorporating an
    optically thick gas \& dust disk. 
    {\it(b)} In our models VERT-RIM and VERT-RIM-ACC, the dust component of
    the outer flared dust disk is truncated at the dust sublimation radius,
    while the gas located at smaller radii is either optically thin (Model
    VERT-RIM, Sect.~\ref{cha:RTDISK}) or optically thick (Model VERT-RIM-ACC,
    Sect.~\ref{cha:RTDISKACC}).
    {\it(c)} Due to direct stellar heating, the scale height at the inner dust
    rim might be increased, resulting in a puffed-up inner rim.  Furthermore,
    various effects, such as a pressure-dependent dust evaporation temperature
    and dust segregation towards the disk midplane, will cause the rim to
    curve.  We investigate the influence of such a curved rim shape in our
    models CURV-RIM (Sect.~\ref{cha:RTDISK}) and CURV-RIM-ACC
    (Sect.~\ref{cha:RTDISKACC}). 
    \label{fig:modelillu}
  }
\end{figure}

Analytical models, both of passive irradiating circumstellar disks
\citep{fri85} as well as viscous, actively accreting disks \citep{lyn74},
predict that the radial temperature profile of YSO disks should follow a
simple power-law $T(r) \propto r^{-\alpha}$.  Most studies infer
a power law index of $\alpha = 3/4$ \citep[][]{mil01, eis05} or $1/2$
\citep[e.g.\ ][]{lei04}.
Using this radial temperature power law and the assumption that each disk
annulus radiates as a blackbody, we can compute the wavelength-dependence of
the disk size corresponding to a certain analytic model.
To model the wavelength-dependence of the size of an accretion disk model with
constant power law, we simulated disks with $T(r) = r^{-\alpha}$, where $r$ is
chosen to be in units of the disk inner radius with a temperature at the inner
dust rim of $T(1)\equiv 1\,500$~K.  The outer disk radius is chosen so that the
temperature drops well below 100~K.  After computing the intensity profile, 
we determine the disk size $a(\lambda)$ for these analytic disk models using
the half-light radius definition by \citet{lei04}. 
Finally, we scale the half-light diameter to fit the NIR (or MIR) size
measured on MWC~147 and show the resulting $a(\lambda)$-curves for
the afore-mentioned representative values for $\alpha$
(Fig.~\ref{fig:diamvariation}).

\begin{figure*}[p]
  \begin{center}
    \vspace{-0.5cm}
    \includegraphics[height=18.0cm]{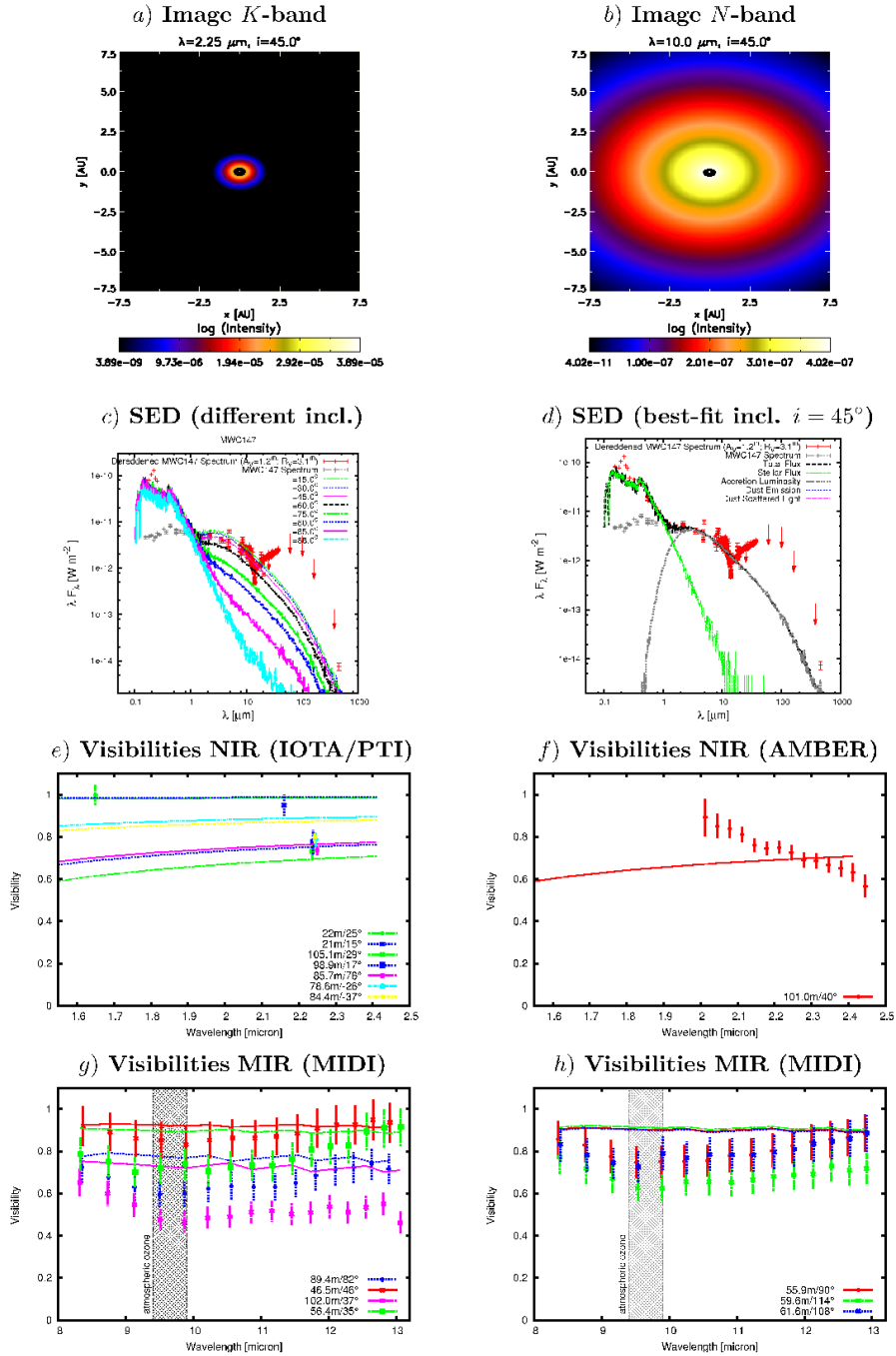}
  \end{center}
  \caption[MWC~147 -- Classical accretion disk model]{
    Analytic accretion disk model, assuming the \citet{hil92} disk
    temperature profile ($\boldmath \chi^2_r=5.56 \boldmath$).
    $a)$ and $b)$ show the corresponding brightness distribution for two
    representative NIR ($2.25~\mu$m) and MIR ($10.0~\mu$m) wavelengths.
    $c)$ shows the SED for various inclination angles, whereas $d)$ gives the
    SED for the best-fit inclination angle and separates the flux which
    originates in stellar photospheric emission and from the accretion disk.
    Finally, $e)$ and $f)$ depict the NIR and MIR visibilities.
    \label{fig:RTmodelHIL92}
  }
\end{figure*}

\citet{hil92} fitted the SEDs of HAeBe stars with
 analytic disk models that combined the
  temperature profile of the reprocessing and actively accreting disk
  component.  This class of ``classical'' optically thick, geometrically
  thin accretion disk models (Fig.~\ref{fig:modelillu}{\it a}) was also
  favoured by various authors to model the SED and NIR broadband visibilities
  of Herbig~Be stars \citep{eis04, mon05, vin07}. \citet{vin07} modeled a
  large set of archival interferometric data and concluded that such classical
  accretion disk models are consistent with the observed visibilities
 of Herbig~Be stars.
  Therefore, we examined whether such models can reproduce our
  combined NIR/MIR spectro-interferometric data on MWC~147.
  In order to fit this analytic model with precisely the same procedure as
  for our radiative transfer modeling (Sect.~\ref{cha:RTfitting}), we computed
  the temperature profile using the equations given in \citet{hil92} and
  included the blackbody emission from an infinitely thin disk in our
  radiative transfer grid.
  Using the stellar parameters given in  Tab.~\ref{tab:stellarproperties},
  we found that a disk model with the same parameters as derived by
 \citet{hil92} for this star (i.e.~inner and outer disk radii of
  $R_{in}=12~R_{\star}$ and $R_{out}=62$~AU) provides a good fit to the
  observed SED for an intermediate disk inclination angle
  ({45\degr}, see Sect.~\ref{cha:resgeomodelfitelong}) and assuming
  a mass accretion rate of
  $\dot{M}_{\rm acc} \approx 4\times10^{-5}~M_{\sun}$yr$^{-1}$.
  As shown in Fig.~\ref{fig:RTmodelHIL92}, this model
  can also reproduce the absolute
  level of the NIR visibilities, but it fails to reproduce the spectral
  dependence of the VLTI/AMBER visibilities and also underestimates the MIR
  size, resulting in a rather high $\chi^2_r$ of 5.56.

It is obvious that these analytic models cannot reproduce the measured 
NIR and MIR-sizes simultaneously.
To understand the strong increase in the apparent size with wavelength,
contributions from the following effects might be of importance:

\textit{(a)} Analytic disk models generally do not include the 
effects of scattered light, which can provide significant heating
of the outer parts of the disk and, thus, 
increase the apparent disk size at MIR wavelengths.

\textit{(b)} It has been suggested that the disks around YSOs may not be flat
but flare with increasing radius.  Such a flaring is expected from
vertical hydrostatic equilibrium considerations \citep{ken87}.  As a result,
the outer disk regions intercept more stellar flux, resulting in an increased
luminosity and apparent size at MIR wavelengths.

\textit{(c)} The flux contribution from an extended cold envelope
$f_\mathrm{env/tot}(\lambda)$ might be non-negligible in the MIR.

\textit{(d)} The NIR size might be underestimated, if the amount of NIR
emission originating from close to the star is inadequately estimated 
(e.g.\ due to a biased $f_\mathrm{star/disk}(\lambda)$ or additional accretion
luminosity).

This enumeration illustrates that the currently routinely applied 
analytic disk models contain several problematic points. A more physical and
consistent approach requires detailed radiative transfer modeling.
In the next section, we will present 2-D radiative transfer modeling for
MWC~147.

\section{2-D Radiative Transfer Simulations} \label{cha:RT}


\begin{deluxetable}{lcccccc}
\tabletypesize{\scriptsize}
\tablecaption{Parameters and fitting results for our 2-D radiative transfer models.
\label{tab:RTparameters}}
\tablewidth{0pt}
\tablehead{
  \colhead{}   & \colhead{}   & \colhead{SHELL} & \colhead{VERT-RIM} & \colhead{CURV-RIM} & \colhead{VERT-RIM-ACC}       & \colhead{CURV-RIM-ACC}\\
  \colhead{}   & \colhead{}   & \colhead{(Fig.~\ref{fig:RTmodelSHELL})}      & \colhead{(Fig.~\ref{fig:RTmodelVRIM})}       & \colhead{(Fig.~\ref{fig:RTmodelCRIM})}              & \colhead{(Fig.~\ref{fig:RTmodelVRIMACC})} & \colhead{(Fig.~\ref{fig:RTmodelCRIMACC})}
}
\startdata
  Outer radius                              & [AU]                 & 100                  & 100                  & 100             & 100                  & 100       \\
  Dust density at 10~AU\tablenotemark{a}    & [g$\cdot$cm$^{-3}$]  & $2.8\times 10^{-21}$  & $1.1\times 10^{-18}$ & $1.6\times 10^{-18}$ & $9.4\times 10^{-19}$ & $8.0\times 10^{-19}$ \\
  Grain size distribution\tablenotemark{b}  &                      & Small                & Small \& Large       & Small \& Large  & Small \& Large       & Small \& Large\\
  Radial power law exponent $p$             &                      & 3/2                  & 15/8                 & 15/8            & 15/8                 & 15/8      \\
  Vertical power law exponent $q$           &                      & --                   & 9/8                  & 9/8             & 9/8                  & 9/8       \\
  Mass accretion rate\tablenotemark{c}      & [$M_{\sun}$yr$^{-1}$] & --                   & --                   & --              & $7.0 \times 10^{-6}$  & $6.5 \times 10^{-6}$ \\
  \tableline
  Inclination                               &                      & --                   & 50\degr              & 40\degr         & 50\degr              & 40\degr   \\
  Best fit PA                               &                      & --                   & 50\degr              & 30\degr         & 80\degr              & 110\degr   \\
  $\chi^2_r$ NIR                            &                      & 316.7                & 135.2                & 119.4           & 2.07                 & 2.71      \\
  $\chi^2_r$ MIR                            &                      & 33.8                 & 3.72                 & 1.60            & 0.79                 & 0.95      \\
  \tableline
  $\chi^2_r$ total                          &                      & 80.2                 & 25.3                 & 20.9            & 0.99                 & 1.24      \\
\enddata
\tablecomments{
  All models include an \textbf{extended outer
  envelope} in order to fit the MIR to FIR SED (see
  Sect.~\ref{cha:RTfitting}).  The envelope is composed of small dust
  grains, with a density distribution $\rho~\propto~r^{-1/2}$ with 
  $\rho_{0} = 3 \times 10^{-23}$~g$\cdot$cm$^{-3}$ at $r_{0}=10$~AU
  (see equation \ref{eqn:mwc147envelope}).
  The outer cutoff radius $r_{\rm cutoff} = 40\,000$~AU was chosen such that the dust temperature 
  drops below 10~K.\\
  For all models, we assume a \textbf{minimum dust density} of $10^{-25}$~g$\cdot$cm$^{-3}$.\\
}
\tablenotetext{a}{{\bf Dust Density}, as measured at a radius of 10~AU in the disk midplane.}
\tablenotetext{b}{{\bf Dust Composition~--~}Mixture of warm silicates
  \citep{oss92} and amorphous carbon \citep{han88} (to equal parts).
The size distribution follows $n(a)~\propto~a^{-3.5}$.  For {\it Small Grains} we choose $a_{\rm min}=0.005~\mu$m and $a_{\rm max}=1.0~\mu$m, whereas for {\it Large Grains} we use $a_{\rm min}=1.0~\mu$m and $a_{\rm max}=1\,000~\mu$m.}
\tablenotetext{c}{{\bf Inner Gaseous Disk~--~}The inner gaseous disk component is modeled to
  be geometrically thin, optically thick, and to extend from the co-rotation
  radius $R_\mathrm{corot}$ to the dust sublimation radius (see Sect.~\ref{cha:RTDISKACC}).
}
\end{deluxetable}


\begin{figure*}[p]
  \begin{center}
    \vspace{-0.5cm}
    \includegraphics[height=17.0cm]{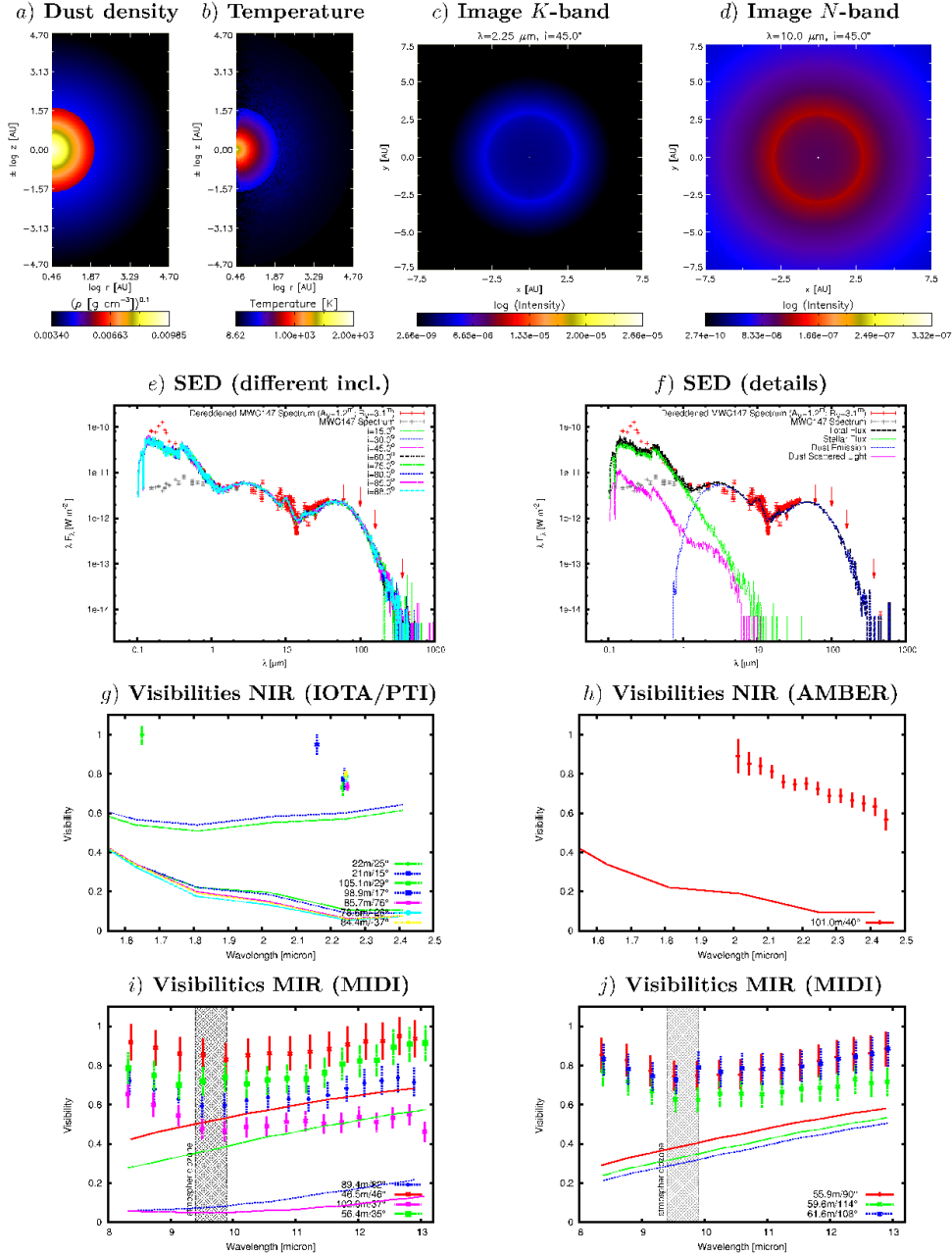}
  \end{center}
  \caption[MWC~147 -- Radiative transfer models - Spherical shell]{
    Radiative transfer model computed for MWC~147 assuming a {\bf Spherical
    Shell geometry} (Model SHELL, $\boldmath \chi^2_r=80.2 \boldmath$, see
    Sect.~\ref{cha:RTSHELL} for a model description). For the model
    parameters, see Tab.~\ref{tab:RTparameters}.  In $a)$ and $b)$, we show
    the dust density and the temperature distribution. $c)$
    and $d)$ show the ray-traced images for two representative NIR
    ($2.25~\mu$m) and MIR ($10.0~\mu$m) wavelengths.
    $e)$ shows the SED for
    various inclination angles, whereas $f)$ gives the SED for the best-fit
    inclination angle and separates the flux which originates in stellar
    photospheric emission, thermal emission, dust irradiation, and accretion
    luminosity. Finally, $g)$ and $h)$ depict the NIR and MIR visibilities
    computed from our radiative transfer models.
    \label{fig:RTmodelSHELL}
  }
\end{figure*}

\begin{figure*}[p]
  \begin{center}
    \vspace{-0.5cm}
    \includegraphics[height=19cm]{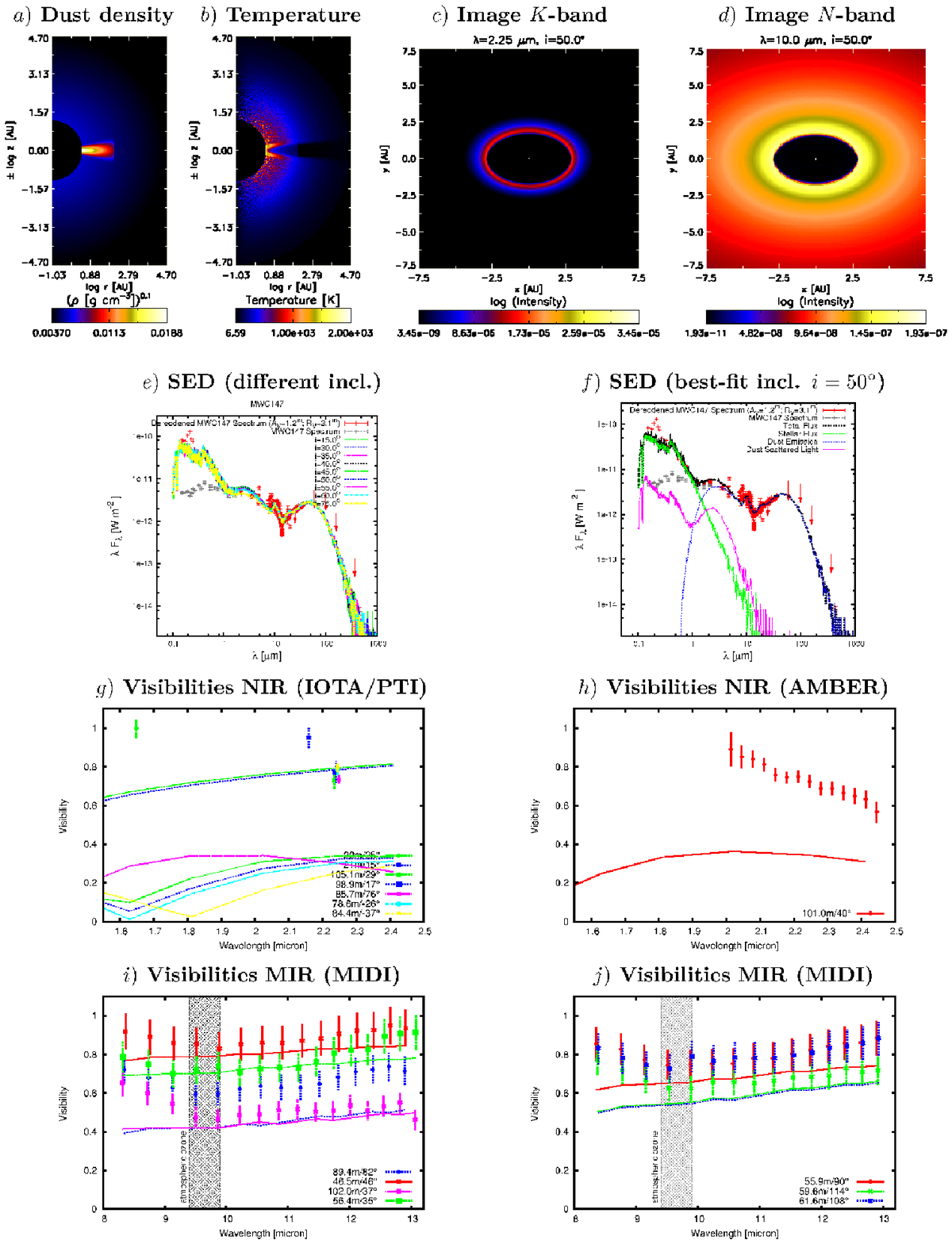}
  \end{center}
  \caption[MWC~147 -- Radiative transfer models - Flared Keplerian
  disk with vertical rim]{Similar to Fig.~\ref{fig:RTmodelSHELL}, but showing the radiative
    transfer model for a {\bf Flared Keplerian disk geometry with a vertical
      inner rim} (Model VERT-RIM, $\boldmath \chi^2_r=25.3 \boldmath$, see
    Sect.~\ref{cha:RTDISK} 
    for a model description).
    \label{fig:RTmodelVRIM}
  }
\end{figure*}

\begin{figure*}[p]
  \begin{center}
    \vspace{-0.5cm}
    \includegraphics[height=19cm]{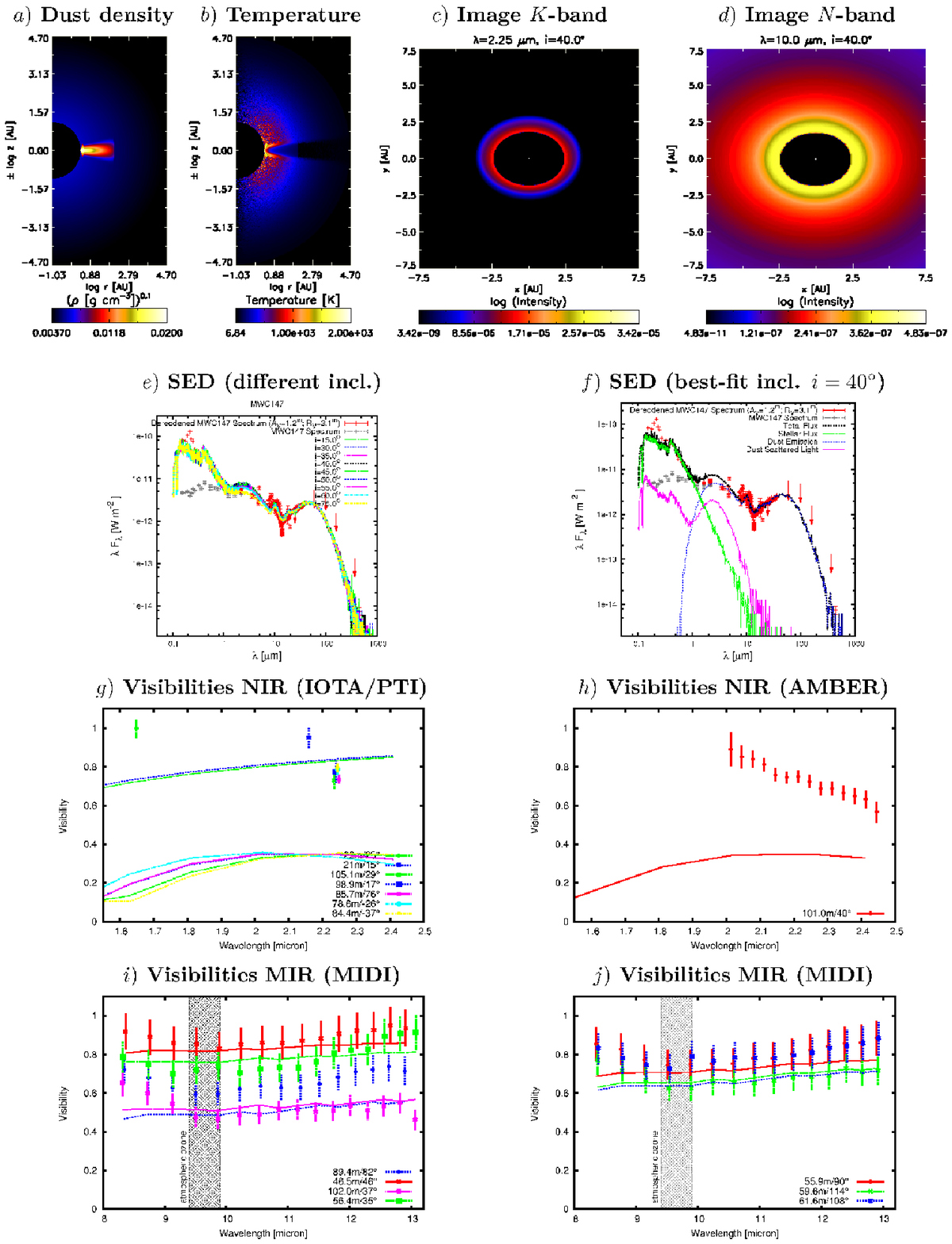}
  \end{center}
  \caption[MWC~147 -- Radiative transfer models - Flared Keplerian
  disk with a curved puffed-up rim]{Similar to Fig.~\ref{fig:RTmodelSHELL},
    but showing the radiative transfer model for a {\bf Flared Keplerian disk
      geometry with a curved puffed-up inner rim} (Model CURV-RIM, $\boldmath \chi^2_r=20.9
    \boldmath$, see Sect.~\ref{cha:RTDISK} for a model description).
    \label{fig:RTmodelCRIM}
  }
\end{figure*}

\begin{figure*}[p]
  \begin{center}
    \vspace{-0.5cm}
    \includegraphics[height=19cm]{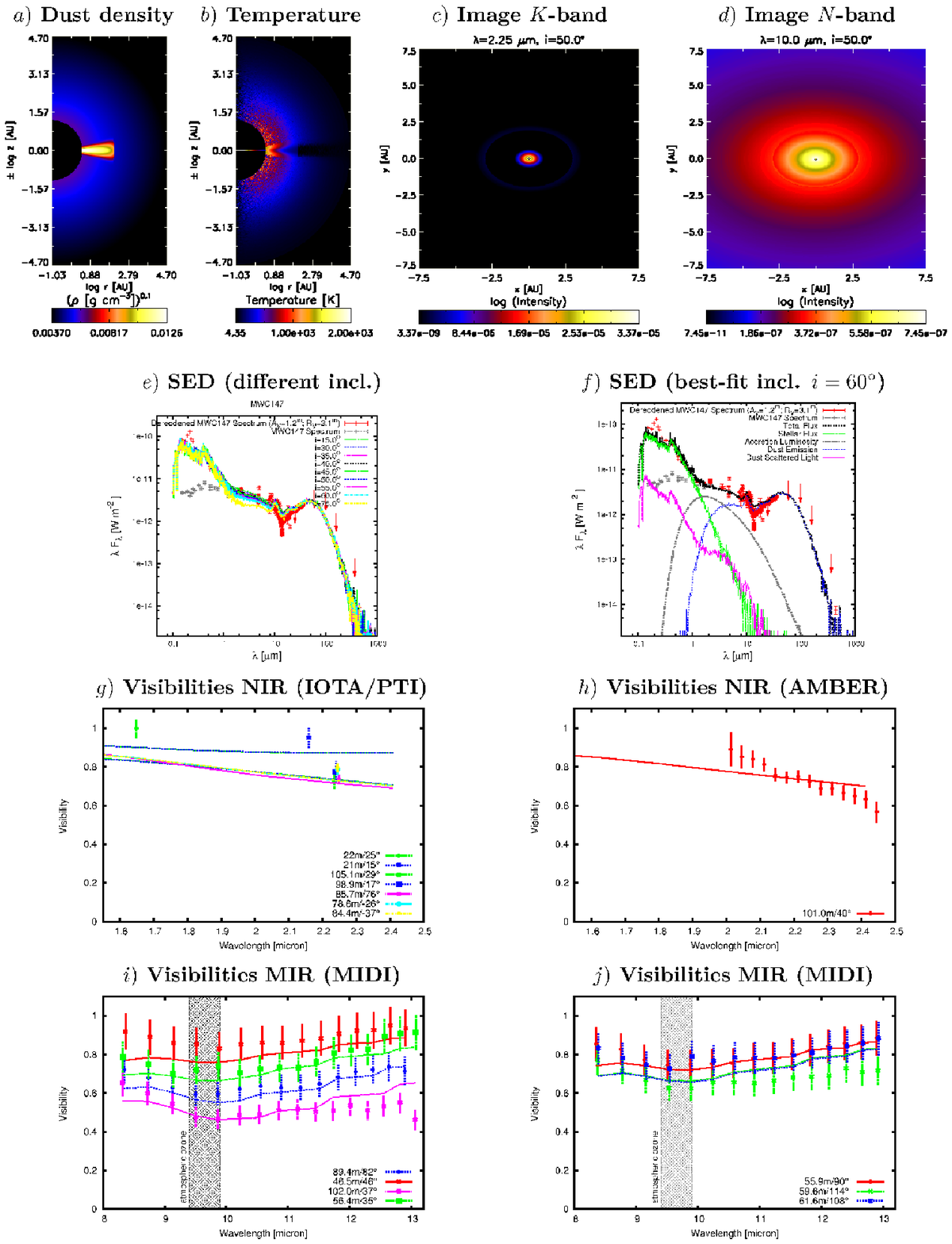}
  \end{center}
  \caption[MWC~147 -- Radiative transfer models - Flared Keplerian
  disk \& Inner Gaseous disk]{Similar to Fig.~\ref{fig:RTmodelSHELL}, but showing the radiative
    transfer model for a {\bf Flared Keplerian disk geometry with a vertical
      inner rim, including accretion} (Model VERT-RIM-ACC, $\boldmath \chi^2_r=0.99 \boldmath$,
    see Sect.~\ref{cha:RTDISKACC} for a model description).
    \label{fig:RTmodelVRIMACC}
  }
\end{figure*}

\begin{figure*}[p]
  \begin{center}
    \vspace{-0.5cm}
    \includegraphics[height=19cm]{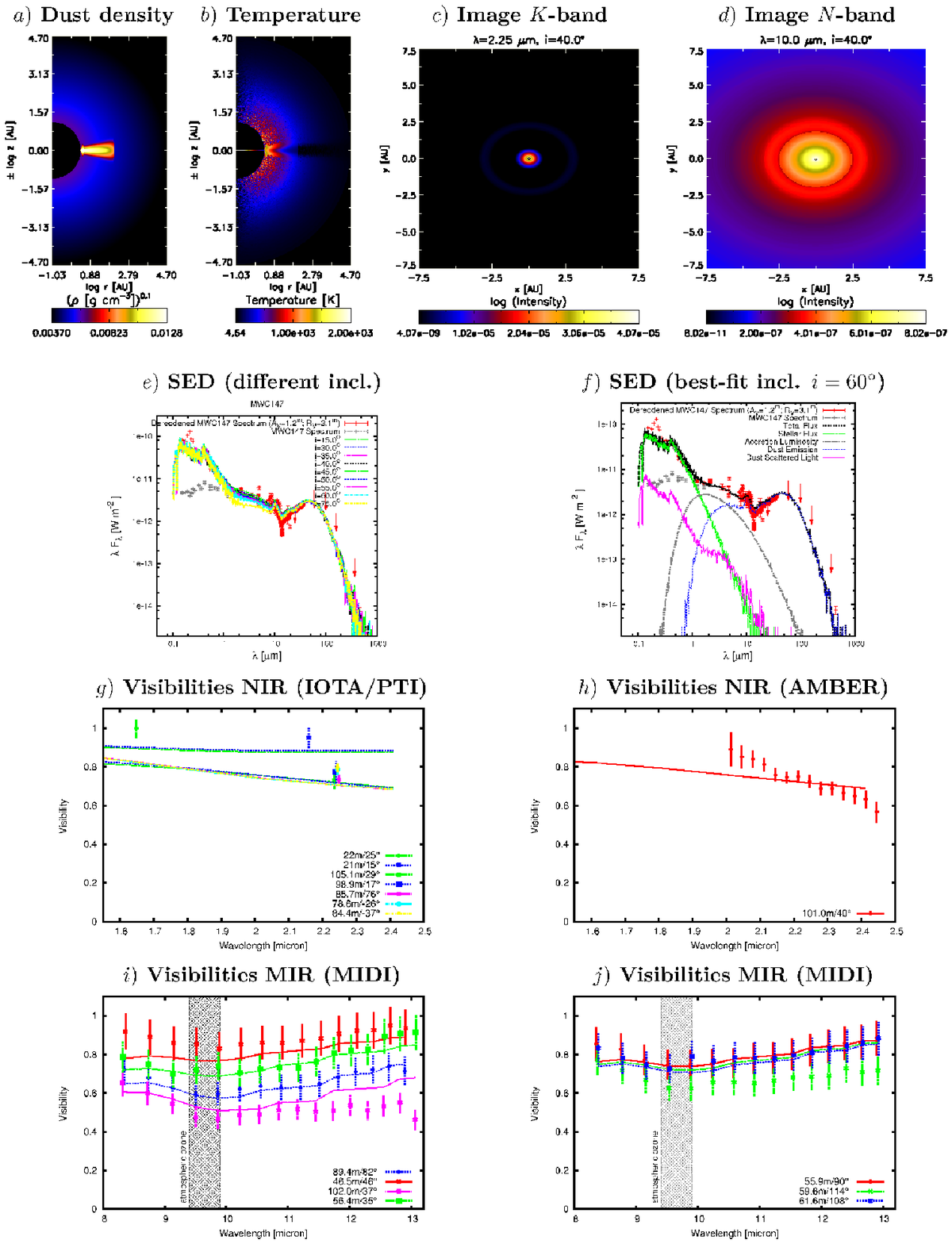}
  \end{center}
  \caption[MWC~147 -- Radiative transfer models - Flared Keplerian
  disk with a curved inner rim \& Inner Gaseous disk]{Similar to 
    Fig.~\ref{fig:RTmodelSHELL}, but showing the radiative transfer model for
    a {\bf Flared Keplerian disk geometry with a curved puffed-up inner rim,
      including accretion} (Model CURV-RIM-ACC, $\boldmath \chi^2_r=1.24 \boldmath$,
    see Sect.~\ref{cha:RTDISKACC} for a model description).
    \label{fig:RTmodelCRIMACC}
  }
\end{figure*}


\subsection{Modeling procedure}

\subsubsection{Monte Carlo radiative transfer simulations} \label{cha:RTcode}

For a physical modeling of our interferometric data, we
employ the radiative transfer code \textit{mcsim\_mpi} (author: K.~Ohnaka),
which solves the radiative transfer problem self-consistently using a Monte
Carlo approach. This code is described in~\citet{ohn06} and has been applied
for the interpretation of interferometric data in \citet{ohn06} and
\citet{hoe06}. 
In short, the stellar flux is treated as a finite number of photon packets 
which are emitted in arbitrary directions.  While propagating through the
cells of the simulation grid, the photon packet can be either scattered or
absorbed.  The probability of these events is given by the density and
the optical properties of the dust in each particular cell.  
For scattering events, the propagation direction of the photon packet changes
anisotropically, according to the Henyey-Greenstein function (this allows us to
also properly treat scattering on large grains).
For each absorption event, energy is deposited into the cell while the packet
is isotropically re-emitted immediately.  The temperature of the
cell is corrected using the scheme by \citet{bjo01}, resulting in a
self-consistent determination of the dust temperature distribution.  
After tracing the propagation of a large number of photon packets through the
simulation grid, the SED is computed by summing the flux from all packets.
The code is parallelized using the LAM/MPI library, which allows the user to
distribute the Monte Carlo computation on a large number of computers within
a network.
We extended the original \textit{mcsim\_mpi} code by
adding an option to include the thermal emission from optically thick
gas (see Sect.~\ref{cha:RTDISKACC}).

As we require particularly high spatial resolution in the inner disk region to
properly resolve the structure of the inner dust rim (at scales of a few AU)
but also need to include structures with large radial extension
(10\,000~AU scale), we employed a spherical grid with logarithmic radial grid
spacing.  The number of radial cells was chosen to be 500, while the
latitudinal grid resolution is 1\degr. We used $5 \times 10^7$ photon packets
per simulation, which ensures sufficient statistics for Monte Carlo radiative
transfer.

As the input stellar spectrum, we use a Kurucz stellar atmosphere
model~\citep{kur70} for a B6-type star of solar metallicity 
($T_\mathrm{eff}=14\,000$~K, $\log g=+4.04$).
For the optical dust properties we use a mixture of warm silicate
\citep{oss92} and amorphous carbon \citep{han88} grains.
The grain size distribution follows the dependence suggested by
\citet[][ $n(a)~\propto~a^{-3.5}$, where $a_{\rm min} \leq a \leq a_{\rm
  max}$]{mat77}.  For the outer envelope, we use small grains 
($a_{\rm min}=0.005~\mu$m; $a_{\rm max}=1.0~\mu$m), whereas for the inner disk
region, we mix in larger dust grains ($a_{\rm min}=1.0~\mu$m; $a_{\rm
  max}=1\,000~\mu$m).
The species of large and small grains are treated separately in the radiative
transfer computations.  A minimum dust density of $10^{-25}$~g$\cdot$cm$^{-3}$ is
assumed.

In order to avoid unrealistically sharp cutoffs at the outer edges of
the considered density distributions, we used a Fermi-type function
$F(r_\mathrm{cutoff})$ to obtain a smooth truncation of the density
distribution around the radius $r_\mathrm{cutoff}$:
\begin{equation}
  F(r_\mathrm{cutoff}) = \left[ 1 + \exp\left(\frac{r-r_\mathrm{cutoff}}{\epsilon \cdot r_\mathrm{cutoff}}\right) \right]^{-1},
  \label{eqn:fermifunction}
\end{equation}
where $\epsilon=0.05$ defines the relative width of the transition region.

\subsubsection{Joint fitting of the SED and wavelength-dependent
  visibilities}\label{cha:RTfitting}

Once the radiative transfer computation is completed, a ray-trace program
is used to compute the SED and synthetic images for any wavelength of interest.
For this project, we compute synthetic images at 3 wavelengths in 
the $H$-band, 3 wavelengths in the $K$-band, and 8 wavelengths in the $N$-band.
Finally,  visibilities are computed from the simulated images for the points
of the $uv$-plane covered by the data. 
In order to treat the visibility slope in the spectro-interferometric
observations properly, we compute the visibility for each spectral channel of
the MIDI and AMBER observations separately, using the synthetic image computed
for a wavelength as close as possible to the central wavelength of the
spectral bin. 

Besides the interferometric observables, we use the SED of MWC~147 
(as shown in FigThe.~\ref{fig:SED}, including the {\it Spitzer}-IRS
spectrum and photometric data from the literature) to constrain
our radiative transfer models.
The contribution of the visual companion \citep[ separation
3\farcs1]{bai06} to the SED is unknown. Furthermore, the photometric
data may be contaminated by the ambient reflection nebula NGC~2247
\citep{cas91}.
\citet{pol02} studied the appearance of the circumstellar environment of
MWC~147 at MIR wavelengths and found an extended structure of $\sim
12$\arcsec~diameter ($\sim 9\,600$~AU) at 10~$\mu$m. 

In the course of our modeling of the SED of MWC~147, we found that 
disk models are not able to reproduce the observed SED long-ward of
$15\,\mu$m (see Fig.~\ref{fig:SED}). 
The large observed fluxes at mid- and far-infrared (FIR) 
wavelengths require the presence of an extended envelope.
We tried various geometries for this envelope, including rotating,
infalling envelopes with and without polar outflow cavities, as used by
\citet{whi03}.
However, the best agreement was obtained for a simple power-law
dust distribution with a rather flat slope
\begin{equation}
  \rho_{\mathrm{env}}(r) = \rho_0 \left( \frac{r}{r_0} \right)^{-1/2} \times
  F(r_\mathrm{cutoff}), \label{eqn:mwc147envelope}
\end{equation}
where $\rho_0$ is the dust density at some characteristic radius $r_0$.

The fitting of the SED and the visibilities was done in the following way:
First, we fixed the geometry of the extended envelope to match the MIR to FIR
SED.  Then, we varied the geometry for the inner component
(Sect.~\ref{cha:RTSHELL} to \ref{cha:RTDISK}) to obtain the best fit to the
NIR and MIR SED.
The SED could be reproduced well with any of the geometrical models considered
below, again demonstrating the highly ambiguous nature of SED fits.
Significant deviations between the model SEDs and the observed SED are
only apparent in the UV spectral range, between $\sim 150$~nm and $\sim 300$~nm.
The observed UV excess is related to accretion activity and originates
probably from the shocks where accreted material crashes onto the
stellar surface. As these accretion shocks are not included in our modeling,
we ignore the SED deviations in the UV spectral range.

Finally, the agreement of the model with the interferometric observables was
tested. The analytic model fits presented in Tab.~\ref{tab:sizesincl} indicate
that the disk should be notably inclined ($\sim 40$ to $60^\circ$); therefore,
we computed each radiative transfer model for different inclinations (from
edge-on to face-on with an interval of $15^\circ$) in order to find the
best agreement both with the spectral shape of the SED and the
visibilities. In total, about one thousand radiative transfer models have been
computed to identify the best-fit models presented in 
Sect.~\ref{cha:RTSHELL} to \ref{cha:RTDISKACC}.

\subsection{Model~SHELL: Spherical shell geometry} \label{cha:RTSHELL}

\citet{mir97} proposed that optically thin shells can reproduce the SED of
HAeBe stars.  For our spherical shell model we assumed a density distribution
given by
\begin{equation}
  \rho_{\mathrm{shell}}(r) = \rho_0 \left( \frac{r}{r_0} \right)^{-p} \times F(r_\mathrm{cutoff}).
\end{equation}
with $p=3/2$.
As can be seen in Fig.~\ref{fig:RTmodelSHELL}, the shell model can reproduce
the measured SED quite well, but completely fails to reproduce 
the NIR and MIR visibilities, 
both the absolute level of the visibilities as well as their
spectral dependence.  Due to this very poor fit to the interferometric
observables ($\chi^2_r=80.2$), we can reject this geometry.  In addition,
spherical models cannot reproduce the elongation revealed by the analytic model
fits (see Sect.~\ref{cha:resgeomodelfit}).

\subsection{Models~VERT-RIM \& CURV-RIM: Passive dust disks} \label{cha:RTDISK}
 
One solution for a passive disk which is gravitationally dominated by the
central star, vertically isothermal, and in vertical hydrostatic equilibrium
is given by a Keplerian disk geometry \citep{sha73} with a density
distribution
\begin{eqnarray}
  \rho_{\mathrm{disk}}(r, z) & = & \rho_{0} \left( \frac{r}{r_0} \right)^{-p} \exp\left[-\frac{\pi}{4} \left( \frac{z}{h_z} \right)^2 \right] \times F(r_\mathrm{cutoff}) \label{eqn:rhodisk}\\ 
  h_z(r) & = & h_{0} r_{0} \left( \frac{r}{r_{0}} \right)^{q},
\end{eqnarray}
where $r$ is the radial distance in the mid-plane, $z$ is the height above the
disk mid-plane, and $h_z$ is the vertical pressure scale height. $h_{0}$ is
the relative geometrical thickness of the disk at the characteristic radius $r_{0}$. 
To reproduce the SED of MWC~147, we find best agreement with a slightly flared
disk geometry \citep[$q=9/8$, ][]{ken87} and with a radial density power law
exponent $p=15/8$ (corresponding to a surface density law $\Sigma(r) \propto r^{-3/4}$).

In the model VERT-RIM, the inner disk edge of the density distribution is
simply truncated, resulting in a vertical wall at the dust sublimation radius
(Fig.~\ref{fig:modelillu}{\it b}).
\citet{nat01} and \citet{dul01} proposed a modification of the disk geometry;
namely, a ``puffed-up'' inner rim.  As the dust at the inner rim (at the dust
sublimation radius) is directly exposed to the stellar radiation, the scale
height in this region will be significantly increased.
\citet{ise05} pointed out that $T_\mathrm{subl}$ depends on the gas density,
which results in a higher sublimation temperature in the disk 
midplane than in the disk atmosphere, causing a curved rim shape.
Recently, \citet{tan07} investigated also the effect of dust segregation on
the rim shape, considering two dust species with different grain sizes which
are distributed over different scale heights.
Both the pressure-dependent dust sublimation temperature, as well as the dust
settling effects, result in a curved shape of the inner dust rim.
In order to investigate the influence of this curved rim on our model fit, we
use the analytic dust segregation model from \citet{tan07} to
compute the dust scale height $h_z$ at the evaporation front as a function of
distance along the disk midplane.  We use the Planck mean opacities
corresponding to the two dust species descripted in Sect.~\ref{cha:RTcode},
and assume that larger grain species has settled to 60\% of the scale height
of the smaller grain population \citep{tan07}. The computed scale height for
the curved rim is combined with the scale height of the outer flared disk,
providing the input density distribution for our radiative transfer simulation
of a disk model with curved inner rim (Model CURV-RIM,
Fig.~\ref{fig:modelillu}{\it c}).

\medskip

For all models with irradiated disks, we find that especially the NIR model
 visibilities are always much smaller than the measured visibilities (see
Fig.~\ref{fig:RTmodelVRIM} and~\ref{fig:RTmodelCRIM}). 
A similar, although less pronounced, deviation was found in the
MIR visibilities.
Therefore, the radiative transfer modeling confirms and quantifies the general
tendency already observed in the geometric model fits; namely, that without
considering accretion luminosity, 
the measured NIR radius of $\sim$0.7~AU (see Tab.~\ref{tab:sizes})
is \textit{a factor of 4 smaller than the dust sublimation radius}.
We conclude that although passive irradiated circumstellar disk models are
able to reproduce the SED of MWC~147, these models are in strong conflict with
the interferometric measurements (resulting in $\chi^2_r=25.3$ for
Model VERT-RIM and $\chi^2_r=20.9$ for Model CURV-RIM; 
see Tab.~\ref{tab:RTparameters} and Fig.~\ref{fig:RTmodelVRIM} and \ref{fig:RTmodelCRIM}).
We would like to point out that adding even larger dust grains cannot solve
this discrepancy, since for grain sizes $\gtrsim 1.2~\mu$m the inner rim
location becomes practically independent of the grain size distribution
\citep[see discussion in][]{ise06}.

\subsection{Models~VERT-RIM-ACC \& CURV-RIM-ACC: Dust disks with active inner
  gaseous accretion disk} \label{cha:RTDISKACC}

In passive circumstellar disks, the infrared emission is generally assumed to
originate almost entirely from dust; the emissivity of the inner dust-free
gaseous part of the disk, at radii smaller than the dust sublimation radius,
is negligible. In an actively accreting disk, on the other hand, viscous
dissipation of energy in the inner dust-free gaseous part of the accretion
disk can heat the gas to high temperatures and give rise to significant
amounts of infrared emission from optically thick gas. The inner edge of this
gas accretion disk is expected to be located a few stellar radii above the
stellar surface, where the hot gas is thought to be channeled towards the star
via magnetospheric accretion columns.
While the magnetospheric accretion columns are too small to be resolved in our
interferometric data (3~$R_{\star}$ correspond to 0.09~AU or 0.12~mas), the
infrared emission from hot gas inside the dust sublimation radius should be
clearly distinguishable from the thermal emission from the dusty disk 
due to the different temperatures of these components and the resulting
characteristic slope in the NIR- and MIR-visibilities.  

As MWC~147 is a quite strong accretor \citep[$\dot{M}_{\rm acc} \approx
10^{-5}~M_{\sun}$yr$^{-1}$, ][]{hil92}, infrared emission from the inner
gaseous accretion disk is likely.
\citet{muz04} found that even for smaller accretion rates the gaseous inner
accretion disk is several times thinner than the puffed-up 
inner dust disk wall and is optically thick (both in radial as well 
as in the vertical direction).

To include the thermal emission from the inner gaseous disk in our radiative
transfer models, we use a similar approach as \citet{ake05b}.
We assume that the accretion luminosity is emitted from a viscous accretion
disk \citep{pri81} which emits at each radius $r$ as a black-body of temperature
\begin{equation}
  T_\mathrm{gas}^4(r) = \left( \frac{3 G M_{\star} \dot{M}}{8 \pi \sigma r^3} \right)
                 \left( 1 - \sqrt{R_{\star}/r} \right)^{1/2}.
\end{equation}
We run $r$ from the magnetic truncation radius $R_\mathrm{corot}$ to the dust
sublimation radius. 
To estimate $R_\mathrm{corot}$ for MWC~147, we use the measured rotation
velocity \citep[$v \sin i = 90$~km\,s$^{-1}$, ][]{boe95}, the stellar
parameters from Tab.~\ref{tab:stellarproperties}, and an intermediate
inclination angle (as derived from the geometric model fits in Tab.~\ref{tab:sizesincl}, 
$i=60^{\circ}$), and yield $R_\mathrm{corot} = G M_{\star} / (2 \pi v^2)
  \approx 3 R_{\star}$.
In our Monte Carlo radiative transfer simulation, the photons
from the gaseous accretion disk are emitted isotropically from the disk
midplane and then propagate through the simulation grid.  In the course of a
simulation, the accretion disk is assumed to be totally optically thick.
Another possible consequence of the presence of an optically thick
inner gas disk may be shielding of the inner dust rim, 
allowing dust to exist closer to the star. However, based on the results of
the theoretical work by \citet{muz04}, we consider this to be a secondary effect and
do not include it in our modeling. 
In order to reproduce the SED in the presence of the excess infrared continuum
emission from optically thick gas, we vary the fraction between small and large
dust grains in the disk in radial direction, which is in qualitative agreement
with the indications for grain growth found in the inner disk regions
(Sect.~\ref{cha:intgraingrowth}).

\medskip

Including the accretion luminosity from an inner gaseous disk improves
the agreement between model predictions and observed visibilities
considerably. 
With a flared disk geometry and an accretion rate of 
$\dot{M}_{\rm acc} = 7 \times 10^{-6}~M_{\sun}$yr$^{-1}$ (see
Tab.~\ref{tab:RTparameters}), both the SED and the interferometric
visibilities (Fig.~\ref{fig:RTmodelVRIMACC} and \ref{fig:RTmodelCRIMACC})
are reasonably well reproduced.
This result is not very sensitive to the precise shape of the inner rim 
(Model VERT-RIM-ACC $\chi^2_r=0.99$; Model CURV-RIM-ACC: $\chi^2_r=1.24$).

\section{Conclusions} \label{cha:conclusions}

We have presented infrared long-baseline interferometric observations of
MWC~147, constraining, for the first time, the inner circumstellar environment
around a Herbig~Be star over the wavelength range from 2 to 13~$\mu$m.

The interferometric data obtained from the PTI archive and with VLTI/AMBER
suggest a characteristic diameter of just $\sim 1.3$~AU (Gaussian FWHM) 
for the NIR emitting region, while the MIR structure is about a factor of 7
more extended (9~AU at 11~$\mu$m). Within the $K$-band, we measure a
significant increase of size with wavelength. 
Comparing the measured wavelength-dependence of the characteristic size 
with the class of analytic accretion disk models, which are currently most
commonly applied for the modeling of HAeBe SEDs \citep{hil92} and NIR interferometric
data \citep[e.g.][]{vin07} yields that these 
models cannot well reproduce the measured significant increase of the
apparent size with wavelength.
To test whether more realistic physical models of the circumstellar dust
environment yield better agreement, we employed 2-D radiative transfer
modeling.  The radiative transfer models were constructed to fit the SED from
0.3 to 450~$\mu$m, including an extended envelope and an inner dust shell/disk
geometry.
While models of passive irradiated disks with or without puffed-up
rims are able to reproduce the SED, they are in conflict with the interferometric
observables, significantly overestimating the size of both the NIR and MIR
emission.
This is also the case, to an even bigger extend, for spherical shell geometries.
Therefore, in our radiative transfer models, we incorporated additional
accretion luminosity emitted from an inner gaseous disk, 
yielding significantly better agreement with the interferometric data. 
The best-fit was obtained with a flared Keplerian disk seen under an inclination
of $\sim 50$\degr, extending out to 100~AU and exhibiting a mass accretion rate of 
$7 \times 10^{-6}~M_{\sun}$yr$^{-1}$.  

Since MWC~147 belongs to the group of ``undersized'' Herbig~Be star disks
(which have measured NIR diameters smaller than expected from the
size-luminosity relation; see \citealt{mon05}), our detailed
spectro-interferometric study confirms earlier speculations that the NIR
emission of Herbig~Be stars in this group might be dominated by gas emission
from inside the dust sublimation radius.

Furthermore, our study demonstrates that the spectro-interferometric capabilities
of the latest generation of long-baseline interferometric instruments are
particularly well suited  to reveal the contributions from active accretion
processes taking place close to the star.  
To achieve further progress on MWC~147, it seems promising for future
spectro-interferometric observations not only to extend the spectral coverage
(e.g.\ to the $J$- and $H$-band) but to employ also a higher spectral
resolution in order to study the spatial origin of accretion-related emission
lines \citep[e.g.\ Br$\gamma$, ][]{gar06}, maybe tracing ionized, optically
thin gas in the disk or magnetospheric accretion \citep{har94}. 
In addition, future long-baseline interferometric observations will measure
the closure phase relation, which is a sensitive measure to asymmetries in 
the source brightness distribution, and which provides also the key to direct 
aperture synthesis imaging of the sub-AU circumstellar environment.

\acknowledgments

We thank Thomas Driebe, who developed software tools which were used in
the course of the reduction of the VLTI/MIDI data. We are also
grateful to Rachel Akeson for advice on the reduction of the PTI data, and
to the anonymous referee, whose detailed referee report helped to improve
this paper.
SK was supported for this research through a fellowship from the International
Max Planck Research School (IMPRS) for Radio and Infrared Astronomy at the 
University of Bonn.

The Palomar Testbed Interferometer is operated by the Michelson Science Center
and the PTI collaboration and was constructed with funds from the Jet
Propulsion Laboratory, Caltech as provided by the National Aeronautics and
Space Administration.
This work has made use of services produced by the Michelson Science Center at
the California Institute of Technology.

This work is based, in part, on archival data obtained with the Spitzer Space
Telescope, which is operated by the Jet Propulsion Laboratory, California
Institute of Technology under a contract with NASA. 

SMART was developed by the IRS Team at Cornell University and is available
through the Spitzer Science Center at Caltech.

This publication makes use of data products from the Two Micron All Sky
Survey, which is a joint project of the University of Massachusetts and the
Infrared Processing and Analysis Center/California Institute of Technology,
funded by the National Aeronautics and Space Administration and the National
Science Foundation.

\bibliographystyle{apj}
\bibliography{ms}

\begin{thebibliography}{}

\bibitem[\protect\citeauthoryear{{Acke} \& {van den Ancker}}{{Acke} \& {van den
  Ancker}}{2004}]{ack04}
{Acke}, B.,  \& {van den Ancker}, M.~E. 2004, \aap, 426, 151

\bibitem[\protect\citeauthoryear{{Akeson} et~al.}{{Akeson}
  et~al.}{2000}]{ake00}
{Akeson}, R.~L., {Ciardi}, D.~R., {van Belle}, G.~T., {Creech-Eakman}, M.~J.,
  \& {Lada}, E.~A. 2000, \apj, 543, 313

\bibitem[\protect\citeauthoryear{{Akeson} et~al.}{{Akeson}
  et~al.}{2005}]{ake05b}
{Akeson}, R.~L., et~al. 2005, \apj, 622, 440

\bibitem[\protect\citeauthoryear{{Allamandola}, {Tielens}, \&
  {Barker}}{{Allamandola} et~al.}{1985}]{all85}
{Allamandola}, L.~J., {Tielens}, A.~G.~G.~M.,  \& {Barker}, J.~R. 1985, \apjl,
  290, L25

\bibitem[\protect\citeauthoryear{{Baines} et~al.}{{Baines}
  et~al.}{2006}]{bai06}
{Baines}, D., {Oudmaijer}, R.~D., {Porter}, J.~M.,  \& {Pozzo}, M. 2006,
  \mnras, 367, 737

\bibitem[\protect\citeauthoryear{{Berrilli} et~al.}{{Berrilli}
  et~al.}{1987}]{ber87}
{Berrilli}, F., {Lorenzetti}, D., {Saraceno}, P.,  \& {Strafella}, F. 1987,
  \mnras, 228, 833

\bibitem[\protect\citeauthoryear{{Bertout}, {Robichon}, \& {Arenou}}{{Bertout}
  et~al.}{1999}]{ber99}
{Bertout}, C., {Robichon}, N.,  \& {Arenou}, F. 1999, \aap, 352, 574

\bibitem[\protect\citeauthoryear{{Bjorkman} \& {Wood}}{{Bjorkman} \&
  {Wood}}{2001}]{bjo01}
{Bjorkman}, J.~E.,  \& {Wood}, K. 2001, \apj, 554, 615

\bibitem[\protect\citeauthoryear{{Boehm} \& {Catala}}{{Boehm} \&
  {Catala}}{1995}]{boe95}
{Boehm}, T.,  \& {Catala}, C. 1995, \aap, 301, 155

\bibitem[\protect\citeauthoryear{{Bouret} et~al.}{{Bouret}
  et~al.}{2003}]{bou03}
{Bouret}, J.-C., {Martin}, C., {Deleuil}, M., {Simon}, T.,  \& {Catala}, C.
  2003, \aap, 410, 175

\bibitem[\protect\citeauthoryear{{Brittain} et~al.}{{Brittain}
  et~al.}{2007}]{bri07}
{Brittain}, S.~D., {Simon}, T., {Najita}, J.~R.,  \& {Rettig}, T.~W. 2007,
  \apj, 659, 685

\bibitem[\protect\citeauthoryear{{Casey}}{{Casey}}{1991}]{cas91}
{Casey}, S.~C. 1991, \apj, 371, 183

\bibitem[\protect\citeauthoryear{{Cohen}}{{Cohen}}{1973}]{coh73a}
{Cohen}, M. 1973, \mnras, 161, 105

\bibitem[\protect\citeauthoryear{{Colavita} et~al.}{{Colavita}
  et~al.}{1999}]{col99}
{Colavita}, M.~M., et~al. 1999, \apj, 510, 505

\bibitem[\protect\citeauthoryear{{Corporon} \& {Lagrange}}{{Corporon} \&
  {Lagrange}}{1999}]{cor99}
{Corporon}, P.,  \& {Lagrange}, A.-M. 1999, \aaps, 136, 429

\bibitem[\protect\citeauthoryear{{Dullemond}, {Dominik}, \&
  {Natta}}{{Dullemond} et~al.}{2001}]{dul01}
{Dullemond}, C.~P., {Dominik}, C.,  \& {Natta}, A. 2001, \apj, 560, 957

\bibitem[\protect\citeauthoryear{{Egan} et~al.}{{Egan} et~al.}{1999}]{ega99}
{Egan}, M.~P., {Price}, S.~D., {Shipman}, R.~F., {Gugliotti}, G.~M., {Tedesco},
  E.~F., {Moshir}, M.,  \& {Cohen}, M. 1999, in ASP Conf. Ser. 177:
  Astrophysics with Infrared Surveys: A Prelude to SIRTF, ed. M.~D. {Bicay},
  R.~M. {Cutri}, \& B.~F. {Madore}, 404

\bibitem[\protect\citeauthoryear{{Egret} et~al.}{{Egret} et~al.}{1992}]{egr92}
{Egret}, D., {Didelon}, P., {McLean}, B.~J., {Russell}, J.~L.,  \& {Turon}, C.
  1992, \aap, 258, 217

\bibitem[\protect\citeauthoryear{{Eisner} et~al.}{{Eisner}
  et~al.}{2007}]{eis07}
{Eisner}, J.~A., {Chiang}, E.~I., {Lane}, B.~F.,  \& {Akeson}, R.~L. 2007,
  \apj, 657, 347

\bibitem[\protect\citeauthoryear{{Eisner} et~al.}{{Eisner}
  et~al.}{2005}]{eis05}
{Eisner}, J.~A., {Hillenbrand}, L.~A., {White}, R.~J., {Akeson}, R.~L.,  \&
  {Sargent}, A.~I. 2005, \apj, 623, 952

\bibitem[\protect\citeauthoryear{{Eisner} et~al.}{{Eisner}
  et~al.}{2004}]{eis04}
{Eisner}, J.~A., {Lane}, B.~F., {Hillenbrand}, L.~A., {Akeson}, R.~L.,  \&
  {Sargent}, A.~I. 2004, \apj, 613, 1049

\bibitem[\protect\citeauthoryear{{Friedjung}}{{Friedjung}}{1985}]{fri85}
{Friedjung}, M. 1985, \aap, 146, 366

\bibitem[\protect\citeauthoryear{{Garcia Lopez} et~al.}{{Garcia Lopez}
  et~al.}{2006}]{gar06}
{Garcia Lopez}, R., {Natta}, A., {Testi}, L.,  \& {Habart}, E. 2006, \aap, 459,
  837

\bibitem[\protect\citeauthoryear{{Habart}, {Natta}, \& {Kr{\"u}gel}}{{Habart}
  et~al.}{2004}]{hab04}
{Habart}, E., {Natta}, A.,  \& {Kr{\"u}gel}, E. 2004, \aap, 427, 179

\bibitem[\protect\citeauthoryear{{Habart} et~al.}{{Habart}
  et~al.}{2006}]{hab06}
{Habart}, E., {Natta}, A., {Testi}, L.,  \& {Carbillet}, M. 2006, \aap, 449,
  1067

\bibitem[\protect\citeauthoryear{{Hanner}}{{Hanner}}{1988}]{han88}
{Hanner}, M. 1988, {Grain optical properties}, Technical report

\bibitem[\protect\citeauthoryear{{Hartmann}, {Hewett}, \& {Calvet}}{{Hartmann}
  et~al.}{1994}]{har94}
{Hartmann}, L., {Hewett}, R.,  \& {Calvet}, N. 1994, \apj, 426, 669

\bibitem[\protect\citeauthoryear{{Helou} \& {Walker}}{{Helou} \&
  {Walker}}{1988}]{hel88}
{Helou}, G.,  \& {Walker}, D.~W., ed. 1988, {Infrared astronomical satellite
  (IRAS) catalogs and atlases. Volume 7: The small scale structure catalog}

\bibitem[\protect\citeauthoryear{{Hern{\'a}ndez} et~al.}{{Hern{\'a}ndez}
  et~al.}{2004}]{her04b}
{Hern{\'a}ndez}, J., {Calvet}, N., {Brice{\~n}o}, C., {Hartmann}, L.,  \&
  {Berlind}, P. 2004, \aj, 127, 1682

\bibitem[\protect\citeauthoryear{{Higdon} et~al.}{{Higdon}
  et~al.}{2004}]{hig04}
{Higdon}, S.~J.~U., et~al. 2004, \pasp, 116, 975

\bibitem[\protect\citeauthoryear{{Hillenbrand} et~al.}{{Hillenbrand}
  et~al.}{1992}]{hil92}
{Hillenbrand}, L.~A., {Strom}, S.~E., {Vrba}, F.~J.,  \& {Keene}, J. 1992,
  \apj, 397, 613

\bibitem[\protect\citeauthoryear{{Hinz}, {Hoffmann}, \& {Hora}}{{Hinz}
  et~al.}{2001}]{hin01}
{Hinz}, P.~M., {Hoffmann}, W.~F.,  \& {Hora}, J.~L. 2001, \apjl, 561, L131

\bibitem[\protect\citeauthoryear{{H{\o}g} et~al.}{{H{\o}g}
  et~al.}{2000}]{hog00}
{H{\o}g}, E., et~al. 2000, \aap, 355, L27

\bibitem[\protect\citeauthoryear{{H{\"o}nig} et~al.}{{H{\"o}nig}
  et~al.}{2006}]{hoe06}
{H{\"o}nig}, S.~F., {Beckert}, T., {Ohnaka}, K.,  \& {Weigelt}, G. 2006, \aap,
  452, 459

\bibitem[\protect\citeauthoryear{{Houck} et~al.}{{Houck} et~al.}{2004}]{hou04}
{Houck}, J.~R., et~al. 2004, \apjs, 154, 18

\bibitem[\protect\citeauthoryear{{Isella} \& {Natta}}{{Isella} \&
  {Natta}}{2005}]{ise05}
{Isella}, A.,  \& {Natta}, A. 2005, \aap, 438, 899

\bibitem[\protect\citeauthoryear{{Isella}, {Testi}, \& {Natta}}{{Isella}
  et~al.}{2006}]{ise06}
{Isella}, A., {Testi}, L.,  \& {Natta}, A. 2006, \aap, 451, 951

\bibitem[\protect\citeauthoryear{{Jaffe}}{{Jaffe}}{2004}]{jaf04}
{Jaffe}, W.~J. 2004, in New Frontiers in Stellar Interferometry, Proceedings of
  SPIE Volume 5491. Edited by Wesley A. Traub. Bellingham, WA: The
  International Society for Optical Engineering, 2004., p.715, ed. W.~A.
  {Traub}, 715

\bibitem[\protect\citeauthoryear{{Jain}, {Bhatt}, \& {Sagar}}{{Jain}
  et~al.}{1990}]{jai90}
{Jain}, S.~K., {Bhatt}, H.~C.,  \& {Sagar}, R. 1990, \aaps, 83, 237

\bibitem[\protect\citeauthoryear{{Kenyon} \& {Hartmann}}{{Kenyon} \&
  {Hartmann}}{1987}]{ken87}
{Kenyon}, S.~J.,  \& {Hartmann}, L. 1987, \apj, 323, 714

\bibitem[\protect\citeauthoryear{{Kraus} et~al.}{{Kraus} et~al.}{2005}]{kra05}
{Kraus}, S., et~al. 2005, \aj, 130, 246

\bibitem[\protect\citeauthoryear{{Kurucz}}{{Kurucz}}{1970}]{kur70}
{Kurucz}, R.~L. 1970, SAO Special Report, 309

\bibitem[\protect\citeauthoryear{{Leinert} et~al.}{{Leinert}
  et~al.}{2004}]{lei04}
{Leinert}, C., et~al. 2004, \aap, 423, 537

\bibitem[\protect\citeauthoryear{{Lynden-Bell} \& {Pringle}}{{Lynden-Bell} \&
  {Pringle}}{1974}]{lyn74}
{Lynden-Bell}, D.,  \& {Pringle}, J.~E. 1974, \mnras, 168, 603

\bibitem[\protect\citeauthoryear{{Malbet} et~al.}{{Malbet}
  et~al.}{2007}]{mal07}
{Malbet}, F., et~al. 2007, \aap, 464, 43

\bibitem[\protect\citeauthoryear{{Malbet} et~al.}{{Malbet}
  et~al.}{1998}]{mal98}
{Malbet}, F., et~al. 1998, \apjl, 507, L149

\bibitem[\protect\citeauthoryear{{Mannings}}{{Mannings}}{1994}]{man94}
{Mannings}, V. 1994, \mnras, 271, 587

\bibitem[\protect\citeauthoryear{{Mathis}, {Rumpl}, \& {Nordsieck}}{{Mathis}
  et~al.}{1977}]{mat77}
{Mathis}, J.~S., {Rumpl}, W.,  \& {Nordsieck}, K.~H. 1977, \apj, 217, 425

\bibitem[\protect\citeauthoryear{{M{\'e}rand}, {Bord{\'e}}, \& {Coud{\'e} Du
  Foresto}}{{M{\'e}rand} et~al.}{2005}]{mer05}
{M{\'e}rand}, A., {Bord{\'e}}, P.,  \& {Coud{\'e} Du Foresto}, V. 2005, \aap,
  433, 1155

\bibitem[\protect\citeauthoryear{{Millan-Gabet}, {Schloerb}, \&
  {Traub}}{{Millan-Gabet} et~al.}{2001}]{mil01}
{Millan-Gabet}, R., {Schloerb}, F.~P.,  \& {Traub}, W.~A. 2001, \apj, 546, 358

\bibitem[\protect\citeauthoryear{{Millan-Gabet} et~al.}{{Millan-Gabet}
  et~al.}{1999}]{mil99}
{Millan-Gabet}, R., {Schloerb}, F.~P., {Traub}, W.~A., {Malbet}, F., {Berger},
  J.~P.,  \& {Bregman}, J.~D. 1999, \apjl, 513, L131

\bibitem[\protect\citeauthoryear{{Min} et~al.}{{Min} et~al.}{2006}]{min06}
{Min}, M., {Dominik}, C., {Hovenier}, J.~W., {de Koter}, A.,  \& {Waters},
  L.~B.~F.~M. 2006, \aap, 445, 1005

\bibitem[\protect\citeauthoryear{{Miroshnichenko} et~al.}{{Miroshnichenko}
  et~al.}{1999}]{mir99}
{Miroshnichenko}, A., {Ivezi{\'c} }, {\v Z}., {Vinkovi{\'c} }, D.,  \&
  {Elitzur}, M. 1999, \apjl, 520, L115

\bibitem[\protect\citeauthoryear{{Miroshnichenko}, {Ivezic}, \&
  {Elitzur}}{{Miroshnichenko} et~al.}{1997}]{mir97}
{Miroshnichenko}, A., {Ivezic}, Z.,  \& {Elitzur}, M. 1997, \apjl, 475, L41

\bibitem[\protect\citeauthoryear{{Monnier} \& {Millan-Gabet}}{{Monnier} \&
  {Millan-Gabet}}{2002}]{mon02}
{Monnier}, J.~D.,  \& {Millan-Gabet}, R. 2002, \apj, 579, 694

\bibitem[\protect\citeauthoryear{{Monnier} et~al.}{{Monnier}
  et~al.}{2005}]{mon05}
{Monnier}, J.~D., et~al. 2005, \apj, 624, 832

\bibitem[\protect\citeauthoryear{{Muzerolle} et~al.}{{Muzerolle}
  et~al.}{2004}]{muz04}
{Muzerolle}, J., {D'Alessio}, P., {Calvet}, N.,  \& {Hartmann}, L. 2004, \apj,
  617, 406

\bibitem[\protect\citeauthoryear{{Natta} \& {Kruegel}}{{Natta} \&
  {Kruegel}}{1995}]{nat95}
{Natta}, A.,  \& {Kruegel}, E. 1995, \aap, 302, 849

\bibitem[\protect\citeauthoryear{{Natta} et~al.}{{Natta} et~al.}{2001}]{nat01}
{Natta}, A., {Prusti}, T., {Neri}, R., {Wooden}, D., {Grinin}, V.~P.,  \&
  {Mannings}, V. 2001, \aap, 371, 186

\bibitem[\protect\citeauthoryear{{Nisini} et~al.}{{Nisini}
  et~al.}{1995}]{nis95}
{Nisini}, B., {Milillo}, A., {Saraceno}, P.,  \& {Vitali}, F. 1995, \aap, 302,
  169

\bibitem[\protect\citeauthoryear{{Ohnaka} et~al.}{{Ohnaka}
  et~al.}{2006}]{ohn06}
{Ohnaka}, K., et~al. 2006, \aap, 445, 1015

\bibitem[\protect\citeauthoryear{{Oliver}, {Masheder}, \& {Thaddeus}}{{Oliver}
  et~al.}{1996}]{oli96}
{Oliver}, R.~J., {Masheder}, M.~R.~W.,  \& {Thaddeus}, P. 1996, \aap, 315, 578

\bibitem[\protect\citeauthoryear{{Ossenkopf}, {Henning}, \&
  {Mathis}}{{Ossenkopf} et~al.}{1992}]{oss92}
{Ossenkopf}, V., {Henning}, T.,  \& {Mathis}, J.~S. 1992, \aap, 261, 567

\bibitem[\protect\citeauthoryear{{Pasinetti Fracassini} et~al.}{{Pasinetti
  Fracassini} et~al.}{2001}]{pas01}
{Pasinetti Fracassini}, L.~E., {Pastori}, L., {Covino}, S.,  \& {Pozzi}, A.
  2001, \aap, 367, 521

\bibitem[\protect\citeauthoryear{{Petrov} et~al.}{{Petrov}
  et~al.}{2007}]{pet07}
{Petrov}, R.~G., et~al. 2007, \aap, 464, 1

\bibitem[\protect\citeauthoryear{{Petrov} et~al.}{{Petrov}
  et~al.}{2003}]{pet03}
{Petrov}, R.~G., et~al. 2003, in Interferometry for Optical Astronomy II.
  Edited by Wesley A. Traub . Proceedings of the SPIE, Volume 4838, pp. 924-933
  (2003)., ed. W.~A. {Traub}, 924

\bibitem[\protect\citeauthoryear{{Polomski} et~al.}{{Polomski}
  et~al.}{2002}]{pol02}
{Polomski}, E.~F., {Telesco}, C.~M., {Pi{\~n}a}, R.,  \& {Schulz}, B. 2002,
  \aj, 124, 2207

\bibitem[\protect\citeauthoryear{{Preibisch} et~al.}{{Preibisch}
  et~al.}{2006}]{pre06}
{Preibisch}, T., {Kraus}, S., {Driebe}, T., {van Boekel}, R.,  \& {Weigelt}, G.
  2006, \aap, 458, 235

\bibitem[\protect\citeauthoryear{{Pringle}}{{Pringle}}{1981}]{pri81}
{Pringle}, J.~E. 1981, \araa, 19, 137

\bibitem[\protect\citeauthoryear{{Przygodda} et~al.}{{Przygodda}
  et~al.}{2003}]{prz03}
{Przygodda}, F., {Chesneau}, O., {Graser}, U., {Leinert}, C.,  \& {Morel}, S.
  2003, \apss, 286, 85

\bibitem[\protect\citeauthoryear{{Rho} et~al.}{{Rho} et~al.}{2006}]{rho06}
{Rho}, J., {Reach}, W.~T., {Lefloch}, B.,  \& {Fazio}, G.~G. 2006, \apj, 643,
  965

\bibitem[\protect\citeauthoryear{{Richichi} \& {Percheron}}{{Richichi} \&
  {Percheron}}{2002}]{ric02}
{Richichi}, A.,  \& {Percheron}, I. 2002, \aap, 386, 492

\bibitem[\protect\citeauthoryear{{Richichi}, {Percheron}, \&
  {Khristoforova}}{{Richichi} et~al.}{2005}]{ric05}
{Richichi}, A., {Percheron}, I.,  \& {Khristoforova}, M. 2005, \aap, 431, 773

\bibitem[\protect\citeauthoryear{{Shakura} \& {Syunyaev}}{{Shakura} \&
  {Syunyaev}}{1973}]{sha73}
{Shakura}, N.~I.,  \& {Syunyaev}, R.~A. 1973, \aap, 24, 337

\bibitem[\protect\citeauthoryear{{Skinner}, {Brown}, \& {Stewart}}{{Skinner}
  et~al.}{1993}]{ski93}
{Skinner}, S.~L., {Brown}, A.,  \& {Stewart}, R.~T. 1993, \apjs, 87, 217

\bibitem[\protect\citeauthoryear{{Skrutskie} et~al.}{{Skrutskie}
  et~al.}{2006}]{skr06}
{Skrutskie}, M.~F., et~al. 2006, \aj, 131, 1163

\bibitem[\protect\citeauthoryear{{Tannirkulam}, {Harries}, \&
  {Monnier}}{{Tannirkulam} et~al.}{2007}]{tan07}
{Tannirkulam}, A., {Harries}, T.~J.,  \& {Monnier}, J.~D. 2007, \apj, 661, 374

\bibitem[\protect\citeauthoryear{{Tatulli} et~al.}{{Tatulli}
  et~al.}{2006}]{tat06}
{Tatulli}, E., {Millour}, F., {Chelli}, A.,  et~al. 2006, \aap\,accepted

\bibitem[\protect\citeauthoryear{{Thompson} et~al.}{{Thompson}
  et~al.}{1978}]{tho78}
{Thompson}, G.~I., {Nandy}, K., {Jamar}, C., {Monfils}, A., {Houziaux}, L.,
  {Carnochan}, D.~J.,  \& {Wilson}, R. 1978, {Catalogue of stellar ultraviolet
  fluxes. A compilation of absolute stellar fluxes measured by the Sky Survey
  Telescope (S2/68) aboard the ESRO satellite TD-1} (Unknown)

\bibitem[\protect\citeauthoryear{{Turon} et~al.}{{Turon} et~al.}{1993}]{tur93}
{Turon}, C., et~al. 1993, Bulletin d'Information du Centre de Donnees
  Stellaires, 43, 5

\bibitem[\protect\citeauthoryear{{Tuthill} et~al.}{{Tuthill}
  et~al.}{2002}]{tut02}
{Tuthill}, P.~G., {Monnier}, J.~D., {Danchi}, W.~C., {Hale}, D.~D.~S.,  \&
  {Townes}, C.~H. 2002, \apj, 577, 826

\bibitem[\protect\citeauthoryear{{van Boekel} et~al.}{{van Boekel}
  et~al.}{2004a}]{van04a}
{van Boekel}, R., et~al. 2004a, \nat, 432, 479

\bibitem[\protect\citeauthoryear{{van Boekel} et~al.}{{van Boekel}
  et~al.}{2004b}]{van04b}
{van Boekel}, R., {Waters}, L.~B.~F.~M., {Dominik}, C., {Dullemond}, C.~P.,
  {Tielens}, A.~G.~G.~M.,  \& {de Koter}, A. 2004b, \aap, 418, 177

\bibitem[\protect\citeauthoryear{{van den Ancker}}{{van den
  Ancker}}{2005}]{van05}
{van den Ancker}, M.~E. 2005, in High Resolution Infrared Spectroscopy in
  Astronomy, ed. H.~U. {K{\"a}ufl}, R.~{Siebenmorgen}, \& A.~F.~M. {Moorwood},
  309

\bibitem[\protect\citeauthoryear{{van den Ancker}, {de Winter}, \& {Tjin A
  Djie}}{{van den Ancker} et~al.}{1998}]{van98}
{van den Ancker}, M.~E., {de Winter}, D.,  \& {Tjin A Djie}, H.~R.~E. 1998,
  \aap, 330, 145

\bibitem[\protect\citeauthoryear{{van Dishoeck}}{{van Dishoeck}}{2004}]{dis04}
{van Dishoeck}, E.~F. 2004, \araa, 42, 119

\bibitem[\protect\citeauthoryear{{Vieira} \& {Cunha}}{{Vieira} \&
  {Cunha}}{1994}]{vie94}
{Vieira}, S.~L.~A.,  \& {Cunha}, N.~C.~S. 1994, Informational Bulletin on
  Variable Stars, 4090, 1

\bibitem[\protect\citeauthoryear{{Vinkovi{\'c}} et~al.}{{Vinkovi{\'c}}
  et~al.}{2006}]{vin06}
{Vinkovi{\'c}}, D., {Ivezi{\'c}}, {\v Z}., {Jurki{\'c}}, T.,  \& {Elitzur}, M.
  2006, \apj, 636, 348

\bibitem[\protect\citeauthoryear{{Vinkovi{\'c}} \& {Jurki{\'c}}}{{Vinkovi{\'c}}
  \& {Jurki{\'c}}}{2007}]{vin07}
{Vinkovi{\'c}}, D.,  \& {Jurki{\'c}}, T. 2007, \apj, 658, 462

\bibitem[\protect\citeauthoryear{{Wesselius} et~al.}{{Wesselius}
  et~al.}{1982}]{wes82}
{Wesselius}, P.~R., {van Duinen}, R.~J., {de Jonge}, A.~R.~W., {Aalders},
  J.~W.~G., {Luinge}, W.,  \& {Wildeman}, K.~J. 1982, \aaps, 49, 427

\bibitem[\protect\citeauthoryear{{Whitney} et~al.}{{Whitney}
  et~al.}{2003}]{whi03}
{Whitney}, B.~A., {Wood}, K., {Bjorkman}, J.~E.,  \& {Wolff}, M.~J. 2003, \apj,
  591, 1049

\bibitem[\protect\citeauthoryear{{Wilkin} \& {Akeson}}{{Wilkin} \&
  {Akeson}}{2003}]{wil03}
{Wilkin}, F.~P.,  \& {Akeson}, R.~L. 2003, \apss, 286, 145

\end{thebibliography}

\end{document}